\documentclass[reprint, amsmath, amssymb, aps]{revtex4-2}

\usepackage{graphicx}					
\usepackage{bm}							
\usepackage{enumerate}

\makeatletter							
\@ifclasswith{revtex4-2}{draft}{		
	\usepackage[inline]{showlabels}
	
}{}
\makeatother

\newcounter{Liste1Ende}

\newcommand{\nn}{\nonumber}
\newcommand{\gl}{\big{(}}
\newcommand{\gr}{\big{)}}

\newcommand{\qq}[1]{``#1''}

\newcommand{\eq}[1]{(\ref{eq:#1})}

\newcommand{\tr}{\mathrm{tr}}

\newcommand{\vp}{\varphi}

\newcommand{\mnb}{_{\mu\nu}}

\newcommand{\mb}{_{\mu}}

\newcommand{\cD}{\mathcal{D}}
\newcommand{\cR}{\mathcal{R}}

\newcommand{\Abar}{\bar{A}}

\newcommand{\gtil}{\tilde{g}}

\newcommand{\vptil}{\tilde{\vp}}

\newcommand{\mtil}{\tilde{m}}

\newcommand{\Sbar}{\bar{S}}

\newcommand{\cL}{\mathcal{L}}

\newcommand{\Etil}{\widetilde{E}}

\newcommand{\Gak}{\Gamma_{k}}
\newcommand{\Gakt}{\Gamma_{k}^{(2)}}
\newcommand{\Gar}{\Gamma_{R}}
\newcommand{\Gart}{\Gamma_{R}^{(2)}}
\newcommand{\Gaba}{\overline{\Gamma}}

\newcommand{\rhotil}{\tilde{\rho}}

\newcommand{\dt}{\partial_{t}}
\newcommand{\dk}{\partial_{k}}

\begin{document}

\title{\LARGE Simplified functional flow equation}

\author{C. Wetterich}

\affiliation{Institut  f\"ur Theoretische Physik\\
Universit\"at Heidelberg\\
Philosophenweg 16, D-69120 Heidelberg}

\begin{abstract}

We adapt the precise definition of the flowing effective action in
order to obtain a functional flow equation with simple properties close to
physical intuition. The simplified flow equation is invariant under local gauge
transformations and suitable for both euclidean and Minkowski signature and
analytic continuation. The cutoff always removes fluctuations close to zeros of
the inverse full propagator. A formulation of the simplified flow equation in
terms of renormalized scale invariant fields permits direct access to scaling
solutions and associated fixed points. Our setting is based on a particular
choice of cutoff function which depends on the macroscopic fields. Corrections
to the simplified flow equation involve a field-dependent modification of the
cutoff for which we discuss a systematic expansion. Truncated solutions for a
scalar field theory in four dimensions suggest a new fixed point with a
field-dependent coefficient of the kinetic term.

\end{abstract}

\maketitle



Functional flow equations for statistical systems or quantum field
theories~\cite{WIL, WEG, WILKO, POL} have a long history. Originally, they were
derived as series of equivalent actions for which the ultraviolet cutoff is
lowered continuously. This procedure integrates out fluctuations in a stepwise
manner, in analogy to the block spin formalism~\cite{KAD}. A more recent
development has shifted the focus to an average action~\cite{CWAA} for which an
infrared cutoff is introduced in the functional integral, corresponding to an
effective action for averages of fields over a finite instead of infinite
volume. An exact flow equation for the effective average action $\Gak$ is based
on an infrared cutoff term which is quadratic in the fields~\cite{CWEFE, RWGT,
TETW, ELL, MOR}. This cutoff suppresses in the functional integral all
fluctuations with momenta $q^2$ smaller than $k^2$. The quadratic form of the
cutoff term ensures the simple one-loop form of the functional differential
equation which describes the dependence of $\Gak$ on the \qq{renormalization
scale} $k$. Many interesting results in a wide area of fields are based on this
equation~\cite{DCE}.

Despite this success, stumbling blocks for several issues require sometimes a
rather high degree of complexity of computations. The first concerns the
implementation of local gauge symmetries. The exact flow equation in the
background field formalism~\cite{RWGT} involves two types of gauge fields: the
background gauge field $\Abar\mb$ and the macroscopic gauge field $A\mb$. The
effective action is invariant only under simultaneous gauge transformations of
$\Abar\mb$ and $A\mb$. Suitable \qq{background field identities} ensure that
there is only one independent gauge field~\cite{RWGT, EHW, FRW, FDP}.
Nevertheless, it is not straight forward to construct a gauge invariant
effective action involving a single gauge field, from which correlation
functions can be extracted directly by taking functional derivatives. This is
particularly cumbersome for quantum gravity since both the field equations and
the correlation function (\qq{primordial fluctuations}) are derived in practice
from a diffeomorphism invariant quantum effective action for a single metric
field~\cite{Wetterich_2015, Wetterich_2017}.

A second issue concerns the analytic structure of the effective action. In
principle, the flow equation can be applied both for the euclidean signature and
Minkowski signature~\cite{FLO}. Beyond simple truncations it is often difficult
to conceive an infrared cutoff that respects the analytic structure of the
correlation functions~\cite{PARE, FLPR, JB, PARE2}. One would like to employ a
cutoff which allows analytic continuation between euclidean and Minkowski
signature at all stages of the flow.

Finally, the use of the exact flow equation extends to many interesting problems
where large fluctuation effects arise from the neighborhood of non-vanishing
\qq{critical momenta}, as the Fermi surface for strongly correlated fermions. It
is, however, not always easy to device a cutoff that first removes precisely
these ``critical fluctuations'', such that they are only consecutively included
as the renormalization scale $k$ is lowered to zero.

In this note we explore to what extent these issues can be addressed by means of
a simplification of the flow equation. The simplified flow equation aims for a
setting where the effects of fluctuations are included with the presence of a
type of infrared cutoff, similar to the effective average action. The
corresponding flow equation should respect local gauge symmetries, be compatible
with analytic continuation and realize a cutoff that automatically removes the
problematic fluctuations. The prize to pay is that this flow equation may not be
exact. Nevertheless, we relate this simplified flow equation to the exact flow
equation of ref.~\cite{CWEFE} by choosing an appropriate cutoff and omitting
certain contributions.

We propose an optimization of the definition of the flowing action for which the
simplified flow equation can be considered as the starting point of a systematic
expansion. The key ingredient is the dependence of the cutoff function on the
macroscopic fields. This induces corrections to the flow equation beyond the one
loop form of ref.~\cite{CWEFE}. We modify the source term in order to cast the
corrections into a simple form. We compute explicitly the general form of the
corrections to the simplified flow equation. For our definition of the flowing
effective action the exact flow equation can be formulated in an implicit form.
The expansion around the simplified flow equation has the character of a type of
loop expansion for a correction term. The simplified flow equation itself is
non-perturbative. It includes already all one-loop effects exactly. Having a
structure similar to the exact flow equation for the effective average action
the simplified flow equation also sums up higher perturbative loops and can be
used for a full non-perturbative analysis. The loop expansion proposed here only
concerns corrections to the precise shape of the infrared cutoff.

Optimization by the choice of a cutoff term which is quadratic in the
fluctuation field and independent of the macroscopic field has already a long
tradition~\cite{LIT}. We aim here for a more general optimization for which the
cutoff term is allowed to depend on the macroscopic fields~\cite{CWGIFE, CWFSI}.
These macroscopic fields are the expectation values of the microscopic fields
over which one integrates in the functional integral. The price to pay is that
the one-loop form of the exact flow equation is no longer guaranteed to hold. We
propose an optimized flow equation which maintains formally the simple one-loop
form. The propagator in the loop involves, however, field-dependent corrections
to the effective infrared cutoff which are, in turn, given by a loop expansion.

For a general definition of a flowing action $\Gak$ we require that for $k=0$
all fluctuations are included, such that $\Gamma_0$ is the quantum effective
action (or Gibbs free energy). This condition still leaves a large freedom for
the choice of a functional integral that defines $\Gak$ for $k$ different from
zero. For an \qq{optimized definition} one has to fix the optimization goals.
For our purpose we ask that $\Gak$ respects all symmetries of the microscopic
physics. In case of local gauge symmetries it should be a gauge invariant
functional of a single gauge field or metric. We further ask for
\qq{universality} of the cutoff in the sense that it automatically affects the
\qq{critical fluctuations}. Finally, we aim for consistency with the analytic
structure. These goals do not necessarily coincide with the goal of best
quantitative accuracy. As compared to the more standard effective average action
the simplified flowing action may also be conceptually more distant from
intuitive procedures as averaging or coarse graining. Our aim here is
qualitative robustness. For other purposes the effective average action or other
forms of optimization may be more appropriate.

\subsection*{Simplified flow equation}

The simplified flow equation is defined by
\begin{equation}
\label{eq:1A}
k\dk\Gamma_k[\vp]=k^{2n}\tr\Big\{\gl1-\frac12E[\vp]\gr\gl\Gamma_R^{(2)}[\vp]+k^2\gr^{-n}\Big\}\
,
\end{equation}
with \qq{renormalized second functional derivative}
\begin{equation}
\label{eq:1B}
\Gamma_R^{(2)}[\vp]=g^{-1/2}N^{-1}[\vp]\Gak^{(2)}[\vp]\gl N^{T}\gr^{-1}[\vp]\ .
\end{equation}
Here $\vp$ is the macroscopic field, which is a vector in the space of internal
or Lorentz indices, and $\Gamma_k^{(2)}[\vp]$ denotes the second functional
derivative of the flowing action with respect to $\vp$. The matrix $N[\vp]$is a
type of field-dependent wave function renormalization. Correspondingly, the
matrix $E[\vp]$ is a field-dependent anomalous dimension. We take
\begin{equation}
\label{eq:7A}
E[\vp]=-N^{-1}\partial_t\gl NN^T\gr\gl N^T\gr^{-1}[\vp] ,
\end{equation}
such that for a $k$-dependence of $NN^T$ of the form
$N[\vp]N^T[\vp]=A[\vp]k^{-\eta}$ one finds $E[\vp]=\eta$. The field-dependent
factors $N[\vp]$ become important if the kinetic term in $\Gamma_k$ is
multiplied by a field-dependent function $\sim NN^T$, as for a field-dependent
Planck mass in quantum gravity, or the field-dependent \qq{kinetial} for the
suggested new fixed point in scalar quantum field theory. The trace in
eq.~\eqref{eq:1A} amounts to a momentum integration or an integration over
positions. We take $n=3$ for $d=3,4$ and $n=2$ for $d=1,2$, such that the
momentum integral is finite for $\Gamma_k^{(2)}\sim q^2$.

For an arbitrary metric $g\mnb$, with
$g=\det(g\mnb)$, the factor $g^{-1/2}$ takes out the factor $g^{1/2}$ arising
from $\Gamma=\int_x\sqrt{g}L$. In flat Euclidean space one has
$g_{\mu\nu}=\delta_{\mu\nu}$ and $g=1$, while for Minkowski space
$g_{\mu\nu}=\eta_{\mu\nu}$ implies $g=-1$. The factor $\sqrt{g}$ takes care of
factors of $i$. The optimized flow equation~\eq{1A} can therefore be used for
an arbitrary signature of the metric. Analytic continuation can be performed by
analytic continuation of the metric, keeping the coordinates fixed. All factors
of $i$ are accounted for by the metric.

We will motivate later the definition of the simplified flow equation~\eq{1A}
and discuss the choice of $N[\vp]$ for eqs.~\eq{1B},~\eq{7A}. We also will
investigate how the corrections as compared to an exact flow equation can be
minimized. For the moment we first highlight a few attractive properties of
eq.~\eq{1A}.
\begin{enumerate}[(i)]
\item The infrared regulator $k^2$ always cuts off small values of
$\Gar^{(2)}\lesssim k^2$. This can concern small momenta or, in a more general
context, a range of \qq{critical momenta} for which $\Gar^{(2)}$ is small and
the fluctuation effects therefore enhanced. Infrared finiteness of the momentum
integral is guaranteed for $\Gar^{(2)}>-k^2$.
\item Ultraviolet finiteness of the momentum integral can always be achieved by
choosing large enough $n$.
\item Concerning analytic continuation the only poles in the momentum integral
can occur for zero eigenvalues of the matrix $\Gart+k^2$, provided $E[\vp]$
remains regular.
\item Massive particles with $m^2\gg k^2$ decouple effectively, with their
contribution suppressed by $k^2/m^2$.
\setcounter{Liste1Ende}{\value{enumi}}
\end{enumerate}

In case of symmetries the matrices $N$ and $E$ are typically block diagonal for
the different representations. For fermions one has an additional minus sign
such that the trace becomes a supertrace. For the fermionic part one also adapts
the cutoff function such that the matrix $\Gar^{(2)}$ in eq.~\eq{1A} is
replaced by its square. The latter is typically quadratic in momentum, similar
to $\Gar^{(2)}$ for bosons. For applications in condensed matter physics the
cutoff concerns the momenta close to the Fermi surface.

For theories with gauge symmetries an effective projection on the physical
fluctuations is achieved by a \qq{physical gauge fixing}. Contributions from
gauge fluctuations and a regularized Faddeev-Popov determinant can be combined
into a universal measure factor which is independent of the precise form of
$\Gamma_k$~\cite{CWGIFE}. This measure contribution has to be added to the
simplified flow equation~\eq{1A}. With $\Gamma_R^{(2)}$ involving covariant
derivatives the flow equation respects the gauge symmetry, provided one employs
a gauge covariant $N[\vp]$. This points to one more central feature of the flow
equation~\eq{1A}:
\begin{enumerate}[(i)]
\setcounter{enumi}{\value{Liste1Ende}}
\item The flow equation respects all symmetries, including local gauge
symmetries.
\setcounter{Liste1Ende}{\value{enumi}}
\end{enumerate}

As as technical advantage we may state
\begin{enumerate}[(i)]
\setcounter{enumi}{\value{Liste1Ende}}
\item A suitable choice of the field-dependent wave function renormalization
$N[\vp]$ can simplify the form of $\Gart$.
\setcounter{Liste1Ende}{\value{enumi}}
\end{enumerate}
For many examples and a large set of truncations of $\Gamma_k$ this can be used
to bring $\Gart$ to a form involving covariant Laplacians in a simple way. One
typically aims for settings where $\Gart$ is block diagonal and $E$ can be
approximated for each block by an anomalous dimension $\eta_i$. The factors
$N^{-1}$ first partly diagonalize $\Gamma_k^{(2)}$ and subsequently perform a
rescaling.

Finally, the flow equation~\eq{1A} offers one more important technical
advantage:
\begin{enumerate}[(i)]
\setcounter{enumi}{\value{Liste1Ende}}
\item Heat kernel methods can be employed for the evaluation of the flow
generator.
\setcounter{Liste1Ende}{\value{enumi}}
\end{enumerate}
For a flow equation for which the flow generator (r.h.s. of eq.~\eq{1A}) can
be expressed as a trace over a function of the matrix $\Gart$ one can employ the
identity
\begin{equation}
\label{eq:5A}
k\dk\Gamma_k=\tr W\gl\Gart+k^2\gr=\int_0^{\infty}\text{d}z\,W(z)B(z)\ ,
\end{equation}
with
\begin{equation}
\label{eq:5B}
B(z)=\int\frac{\text{d}s}{2\pi i}e^{sz}\tr\exp\Big[-s\gl\Gart+k^2\gr\Big]\ .
\end{equation}
For a simple enough form of $\Gart$, for example in a derivative expansion, one
can use standard heat kernel expressions for the trace in eq.~\eq{5B}. An
example is the computation of the flow of a field-dependent effective Planck
mass in quantum gravity~\cite{CWYA}. For gauge invariant $\Gamma_k$ the second
functional derivative is gauge covariant, which makes the gauge invariance of
$B(z)$ and therefore the flow generator manifest.

An interesting approximation to the simplified flow equation considers a simple
scalar field $\vp$ and approximates $NN^T$ by a field-independent wave function
renormalization $Z$. For this approximation in flat Euclidean space the
simplified functional flow equation reads
\begin{equation}
\label{eq:A1}
k\dk\Gak[\vp]=k^{2n}\,\tr\Big\{ \left(1-\frac\eta2\right) \left(Z^{-1}
\Gak^{(2)}[\vp]+k^2\right)^{-n}\Big\}\ .
\end{equation}
This equation has been obtained before from the proper time flow
equation~\cite{OLE, FP, LI, LIA, FP, SP, BOZ, BSW, ZAP, LIPA1, LIPA2, ZAP2, BR,
DA, AH}. It has also been proposed in~\cite{BLPV} for the Wilsonian action,
which differs conceptually from the flowing effective action.

Our setting is based on a functional integral for the flowing action and we do
not use a regularization of heat kernels for the definition of the flowing
action. The precise connection which leads to the coincidence of
eq.~\eqref{eq:A1} with a proper time flow equation is not known to us. This may
constitute an interesting topic of research. One also may raise the question if
there exist possible connections between the full simplified flow
equation~\eqref{eq:1A} and the corrections to it with a regularization of heat
kernels.

The simplified flow equation can also be compared with the background formalism
for the effective average action~\cite{RWGT}. The background formalism involves
in addition to the macroscopic field $\vp$ a background field $\bar{\vp}$ on
which the cutoff may depend. The flowing action $\Gamma_k[\vp,\bar{\vp}]$ is
then a functional of two fields, for example two gauge fields $A_\mu$ and
$\bar{A}_\mu$. Identifying $\bar{A}= A$ the effective action $\Gamma_k[A] =
\Gamma_k[A,A]$ is gauge invariant if the cutoff function involves covariant
derivatives with respect to the background field $\bar{A}$. The second
functional derivative $\Gamma_k^{(2)}[A,\bar{A}]$ which appears in the exact one
loop form of the flow equation is taken at fixed $\bar{A}$. A commonly used
approximation replaces $\Gamma_k^{(2)}[A,\bar{A}]$ by $\Gamma_k^{(2)}[A]$. This
yields a flow equation with a similar structure as the simplified flow equation.
In fact, the background formalism equals the flowing action discussed in this
note up to a modification of the source term. The structure of the corrections
to the exact flow equation in the background field formalism is
known~\cite{RWGT}. They are rather complicated since two gauge fields $A$ and
$\bar{A}$ are involved. The corrections to the simplified flow equation have a
simpler structure. A cutoff depending on the second functional derivative of an
effective action and its relation to heat kernels has been discussed for the
background field formalism in ref.~\cite{Litim_2002, Litim_2002B}.

In order to get a first impression of the simplified flow equation one may
evaluate the flow of the effective potential according to eq.~\eq{A1}. In
momentum space one has
\begin{equation}
\label{eq:A2}
\Gak^{(2)}(q,q')=\frac{\partial^2\Gak}{\partial\vp(-q)\partial\vp(q')}\ ,
\end{equation}
and $k^2$ stands for $k^2\delta(q,q')$. We use conventions with $\int_
q=\int\text{d}^{d}q(2\pi)^{-d}$, $\delta(q,q')=(2\pi)^d\delta(q-q')$, $\tr
A(q,q')=\int_{q,q'}A(q,q')\delta(q',q)$, with
$\delta^2(q,q')=\Omega\delta(q,q')$ and $\Omega$ the spacetime volume. We may
take the truncation
\begin{equation}
\label{eq:A3}
\Gak^{(2)}(q,q')=Z\gl q^2+m^2\gr\delta(q,q')\ ,
\end{equation}
with $m^2(\vp)$ given by the second derivative of the scalar potential
$m^2=\partial^2 U/\partial\vp^2$. For $d=4$, $n=3$ eq.~\eq{A1} yields
\begin{equation}
\label{eq:A4}
k\dk\Gak=\Omega k^6\left(1-\frac\eta2\right)\int_q\gl q^2+m^2+k^2\gr^{-3}\ .
\end{equation}
For a constant scalar field $\vp$ one has $\Gak=\Omega U$, with $U$ the
effective potential flowing as
\begin{align}
\label{eq:A5}
k\dk U=&k^6\left(1-\frac\eta2\right)\int_q\gl q^2+m^2+k^2\gr^{-3}\nn\\
=&\frac{\left(1-\frac\eta2\right)k^6}{32\pi^2\gl m^2+k^2\gr}\ .
\end{align}
Neglecting $\eta$ this yields the same expression as for the standard flow
equation with the Litim cutoff~\cite{LIT}.

\subsection*{Relation to exact flow equation}

The simplified flow equation is typically not an exact equation. Relating it to
the exact flow equation for the effective average action~\cite{CWEFE} will
produce a ``correction term'' to the flow equation. We will discuss in the
second part of this note to which extent the precise definition of the flowing
effective action can be optimized in order to minimize this correction term.
Part of this optimization is the choice of the cutoff function.

Let us start for bosons with the flow equation $(\partial_t=k\partial_k)$
\begin{equation}
\label{eq:MF1}
\partial_t\Gamma_k=\frac12\tr\Big\{\dt R_k\gl\Gakt+R_k\gr^{-1}\Big\}\ ,
\end{equation}
where
\begin{equation}
\label{eq:MF2}
R_k=g^{1/2}k^2Nr(B)N^T\ ,\quad B=k^{-2}\Gart\ .
\end{equation}
This flow equation would be exact only if $R_k$ is independent of $\vp$. In
general, this is not the case here since both matrices $N$ and $B$ are allowed
to depend on $\vp$. The corrections to eq.~\eq{MF1} for field-dependent $B$
and $N$ will be addressed below. We may also consider eq.~\eq{MF1} as the flow
equation in the background formalism for which the correction term discussed in
ref.~\cite{RWGT} has been omitted. Alternatively, it corresponds to the gauge
invariant flow equation without the correction terms specified in
refs.~\cite{CWGIFE, CWFSI}.

We want to demonstrate that for a suitable choice of the cutoff function $R_k$
eq.~\eq{MF1} yields the simplified flow equation~\eq{1A}. This holds up to
small corrections. We approximate
\begin{equation}
\label{eq:MF3}
\dt B=-2B\ ,\quad \dt r=-2B\frac{\partial r}{\partial B}\ ,
\end{equation}
which neglects the $k$-dependence of $\Gart$ in eq.~\eq{MF2}. In this
approximation one obtains for constant $g$
\begin{align}
\label{eq:MF4}
\dt R_k=&\,2g^{1/2}k^2N\tilde rN^T\ ,\\
\tilde r=&\,r+\frac12N^{-1}\dt Nr+\frac12r\dt N^{T}\gl
N^{T}\gr^{-1}-B\frac{\partial r}{\partial B}\ .\nn
\end{align}
Employing
\begin{equation}
\label{eq:MF5}
\Gakt+R_k=g^{1/2}k^2N(B+r)N^{T}
\end{equation}
yields
\begin{align}
\label{eq:MF6}
\dt\Gamma_k=&\,\tr\Big\{\Big[ r+\frac12N^{-1}\dt Nr+\frac12r\dt N^T\gl
N^{T}\gr^{-1}-B\frac{\partial r}{\partial B}\Big]\nn\\
&\quad\times(B+r)^{-1}\Big\}\nn\\
=&\,\tr\Big\{\Big[\left(1-\frac12E\right) r-B\frac{\partial r}{\partial
B}\Big](B+r)^{-1}\Big\}\ .
\end{align}
The second expression neglects a commutator. Together with terms neglected in
eq.~\eq{MF3} this could be (partly) absorbed by a more elaborate definition
of $E[\vp]$.

Let us take a basis for which $B$ is block-diagonal and approximate in each
block $E$ by a constant $\eta_i$. The contribution of a given block to the flow
equation, with $\gamma_i=(1-\eta_i/2)^{-1}$, reads
\begin{equation}
\label{eq:MF7}
\dt\Gak^{(i)}=\tr^{(i)}\Big\{\left(1-\frac{\eta_i}{2}\right)\left(r_i-\gamma_iB_i\frac{\partial
r_i}{\partial B_i}\right)(B_i+r_i)^{-1}\Big\}\ .
\end{equation}
The functions $r_i(x)$ are chosen such that
\begin{equation}
\label{eq:MF8}
(r-\gamma_i x\partial_xr)(x+r)^{-1}=(x+1)^{-n}\ .
\end{equation}
This yields eq.~\eq{1A} with a corresponding approximation for $E$.

Our choice of the cutoff function $r$ corresponds to a solution of the
differential equation~\eq{MF8} with the boundary condition that $r(x)$
vanishes for $x\to\infty$ and approaches a constant for $x\to0$. We demonstrate
this for $n=3$ and $\gamma=1$, where $r$ has to obey
\begin{equation}
\label{eq:N3}
\frac{\text{d}r}{\text{d}x}=(1+x)^{-3}\big[r\gl3+3x+x^2\gr-1\big]\ .
\end{equation}
The solution for large $x$ is given by
\begin{equation}
\label{eq:N4}
r=\frac{1}{3x^2}\ .
\end{equation}
For small $x$ one has
\begin{equation}
\label{eq:N5}
r=a+(3a-1)x\ ,
\end{equation}
where the \qq{integration constant} $a$ is fixed by solving eq.~\eq{N3} with
boundary condition~\eq{N4}. One may expand
\begin{equation}
\label{eq:N6}
r=\frac{1}{3(1+x)^2}+\sum_{j=3}^{\infty}c_j(1+x)^{-j}\ ,
\end{equation}
with iterative relation for $j\geq4$
\begin{equation}
\label{eq:N7}
c_{j+1}=-\frac{1}{j+2}(c_j+c_{j-1})\ ,
\end{equation}
leading to $c_3=-1/12$, $c_4=-1/20$, $c_5=1/45$ etc., and
\begin{equation}
\label{eq:N8}
a=\frac13+\sum_{j=3}^{\infty}c_j\ .
\end{equation}
One can actually find an exact solution of eq.~\eq{N3},
\begin{equation}
\label{eq:D1}
r=1+\frac1z\left[\exp\left(-z-\frac{z^2}{2}\right)-1\right]\ ,\quad
z=\frac{1}{1+x}\ ,
\end{equation}
with $a=r(x=0)=r(z=1)=\exp(-3/2)$. For $x\to-1$, $z\to\infty$, one obtains
$r\to1$ with $r+x\to(1+x)\exp\{-1/\gl2(1+x^2)\gr\}$. The singular behavior of
the pole of the flow equation for $\Gart\to-k^2$ is related to the singular
behavior of $(B+r)^{-1}$ in the flow equation~\eq{MF6}.

In summary, the simple analytic structure of the simplified flow
equation~\eq{1A} is achieved by the adaption of the cutoff function. We see
no obvious reason why similar solutions of the differential equation~\eq{MF8}
should not exist for $\gamma\neq1$. This possibly extends to the inclusion of
corrections to eq.~\eqref{eq:MF3} or more generally to $r$. For an optimal choice of the cutoff function we
believe that the dominant corrections to the simplified flow equation arises
from the corrections to eq.~\eq{MF1} for field-dependent $R_k$. They add to
$R_k$ in eq.~\eq{MF1} a correction term $Q_k$, as we will establish below in
eq.~\eq{EF15}.

\subsection*{Scale invariant fields and scaling form of flow\\equation}

We may express the flow equation in terms of renormalized scale invariant
fields. In four dimensions the renormalized scale invariant scalar field
$\vptil$ and metric $\gtil\mnb$ are given by
\begin{equation}
\label{eq:A6}
\vptil(x)=\frac{N\vp(x)}{k}\ ,\quad \gtil\mnb(x)=k^2g\mnb(x)\ .
\end{equation}
We employ a scale invariant formulation with fixed coordinates, such that the
scale invariant metric appears even for a flat background geometry according to
\begin{align}
\label{eq:A7}
\Gak=&\int_x\sqrt{g(x)}\cL(x)=k^{-4}\int_x\sqrt{\gtil(x)}\cL(x)\nn\\
=&\int_x\sqrt{\gtil(x)}\widetilde\cL(x)\ .
\end{align}
(Alternatively, one could introduce scale invariant coordinates and momenta).
The flow equation at fixed scale invariant fields $\vptil(x)$, $\gtil\mnb(x)$
obtains from eq.~\eq{1A} by a variable transformation,
\begin{align}
\label{eq:A8}
k\dk\Gamma[\vptil,\gtil\mnb]=&\tr\Big\{\left(1-\frac12\Etil\right)
\left(\Gaba^{(2)}+1\right)^{-n}\Big\}\\
&+\int\Big\{\left(1+\frac\eta2\right)\vptil\frac{\partial\Gamma}{\partial\vptil}
-2\gtil\mnb\frac{\partial\Gamma}{\partial\gtil\mnb}\Big\}\nn\ .
\end{align}
The trace and the integral can be performed in position or momentum space, and
$1$ stands for the unit matrix. The matrix $\eta(\vptil)$ is defined by
\begin{equation}
\label{eq:8A}
k\dk N(\vp)=-\frac12\eta(\vp)N(\vp)\ ,
\end{equation}
and we employ
\begin{equation}
\label{eq:8B}
\Gaba_{ab}^{(2)}=\gtil^{-\frac12}\frac{\partial^2\Gamma}{\partial\vptil_a\partial\vptil_b}\
.
\end{equation}

The matrices $E$ and $\Etil$ in eqs.~\eq{1A} and~\eq{A8} are related by
\begin{equation}
\label{eq:8C}
(1-\frac12E)(1+\Gart/k^{2})^{-n}=(1-\frac12\Etil)(1+\Gaba^{(2)})^{-n}\ .
\end{equation}
This absorbs possible connection terms from $\vp$-dependent $N$ into the
generalized anomalous dimension $\Etil$. For $N$ independent of $\vptil$ the
matrices $E$ and $\Etil$ are identical. For a fixed background geometry one has
$\partial\Gamma/\partial\vptil\sim\sqrt{\gtil}$ and the dependence of
$\partial^2\Gamma/\partial\vptil^2$ on $\sqrt{\gtil}$ provides for the correct
dimensional scaling. Scaling solutions obtain as solutions of the differential
equation obtained by setting $\partial_t\Gamma_k[\vptil,\gtil_{\mu\nu}]=0$.
There may exist different versions of scale invariant fields leading possibly to
further scaling solutions.

\subsection*{Flow of scalar effective potential}

We next demonstrate the practical use of the simplified flow equation by a
computation of the flow of the effective potential of the Higgs doublet in the
standard model of particle physics, and the flow of the kinetial
(field-dependent coefficient of kinetic term) for a scalar field theory. For the
flow of the potential we employ directly the scaling form~\eq{A8} of the flow
equation, which yields a very simple expression from which the flow of the
quartic coupling etc. can be inferred by straightforward differentiation. In
four dimensions and for $n=3$ we evaluate eq.~\eq{A8} for constant $\vptil$
and $\gtil\mnb=k^2\eta\mnb$, whereby $\Gamma=\int_x\sqrt{\gtil}u$, with
$u=U/k^4$ the dimensionless scalar potential. For the example of the Higgs
scalar, $\vptil$ is replaced by a complex doublet corresponding to four real
fields $\vptil_\gamma$, and we consider the invariant
$\rhotil=\frac12\sum_\gamma\vptil_\gamma^2$. One infers from eq.~\eq{A8}
\begin{equation}
\label{eq:A9}
k\dk
u(\rhotil)=\sum_i\frac{2-\eta_i}{64\pi^2(1+\mtil_i^2)}+(2+\eta)\rhotil\partial_{\rhotil}u(\rhotil)-4u\
,
\end{equation}
in accordance with eq.~\eq{A5}.

The sum runs over all particles with
renormalized dimensionless mass term $\mtil_i^2=m_i^2/k^2$. Only particles with
small enough $\mtil_i^2$ contribute to the flow in a given range of $k$. For the
scalar fluctuations there are three Goldstone modes with $\mtil_i^2=u'$, while
the radial mode obeys $\mtil^2=u'+2\rhotil u''$. Here primes denote derivatives
with respect to $\rhotil$. For the contribution from fermion fluctuations one
has an additional minus sign and $\mtil_i^2=y_i^2\rhotil$, with $y_i$ the Yukawa
couplings. (Each Majorana fermion contributes a factor $-2$ such that for a
Dirac fermion there is a factor $-4$.) For the two $W$-Bosons one finds
$\mtil_i^2=g_2^2\rhotil/2$ and for the $Z$-boson
$\mtil_i^2=g_2^2\rhotil/2+3g_1\rhotil/10$, with $g_2^2$ and $g_1^2$ the gauge
couplings of the electroweak gauge groups $SU(2)$ and $U(1)$. For each gauge
boson there is a factor three corresponding to the number of physical degrees of
freedom for a massive gauge boson. The measure factor subtracts for each gauge
boson the contribution of one field with $\mtil_i^2=0$. At the minimum of the
potential this cancels the contribution of the scalar Goldstone modes, such that
only contributions from physical fluctuations remain. The contribution from
physical metric fluctuations can be included in a similar way~\cite{PRWY, ESPA,
CWQGS}. Eq.~\eq{A5} holds for an arbitrary constant metric, and therefore both
for Euclidean or Minkowski flat space.

The flow of $\rhotil$-derivatives of $u$ obtains by differentiating both sides
of eq.~\eq{A9} with respect to $\rhotil$. For the quartic scalar coupling
$\lambda(\rhotil)=u''(\rhotil)$ this yields
\begin{align}
k\dk&\lambda(\rhotil)=k\dk u''(\rhotil)=2\eta u''\nn\\
+\frac{1}{16\pi^2}\bigg\{&\left(\frac{3}{(1+u')^3}+\frac{9}{(1+u'+2\rhotil
u'')^3}\right)(u'')^2\nn\\
&-\frac{12y_t^2}{(1+y_t^2\rhotil)^3}+\frac{3g_2^4}{2(1+g_2^2\rhotil/2)^3}\nn\\
&+\frac{3(g_2^2+3g_1^2/5)^2}{4(1+g_2^2\rhotil/2+3g_1^2\rhotil/10)^3}\bigg\}\ .
\end{align}
Here we have neglected the anomalous dimensions $\eta_i$ in the numerators
$\sim(2-\eta_i)$, as well as $u'''$ and $u^{(4)}$. We only include the top quark
Yukawa coupling $y_t$ and take $\rhotil$-independent $y_t$, $g_2$, $g_1$. We may
evaluate the quartic scalar coupling at the potential minimum ($u'=0$) in a
range of small enough $\rhotil$, such that the mass terms in the denominators
can be neglected. This approximation yields the perturbative one-loop running,
provided we also employ the one-loop value for the anomalous dimension of the
Higgs scalar $\eta=\gl6y_t^2-9g_2^2/2-9g_1^2/10\gr/\gl16\pi^2\gr$. As $k$ is
lowered and $\rhotil$ increased the flow provides for appropriate thresholds for
the decoupling of massive particles. Metric fluctuations are not included since
they play a role only for $k$ in the vicinity of the effective Planck mass.

This rather simple computation demonstrates the power of the simplified flow
equation. The effective potential is a manifestly gauge invariant quantity. In
the approximation used here analytic continuation is trivial. Known one-loop
results are reproduced in a simple way, supplemented by a natural description of
threshold effects due to the decoupling of heavy particles. Extending the
truncation in various directions proceeds by more elaborate approximations of
the effective propagator. Two loop accuracy of the flow
equation~\cite{Papenbrock_1995} presumably needs the correction to the
simplified flow equation discussed below.

\subsection*{Flow of kinetic term and kinetic fixed point}

As a second practical demonstration we compute the flow of the kinetial for the
quantum field theory of a single real scalar field $\vp$. The kinetial $K(\vp)$
is the field-dependent coefficient of the term with two derivatives. This
field-dependence induces interactions which can lead to complex features if $K$
is not close to a constant~\cite{wetterich2024field}. For $K\sim\vp^{-2}$ the
model exhibits scale symmetry with respect to multiplicative rescaling of $\vp$.
This symmetry requires a flat potential. One would like to implement a flow
equation respecting this symmetry in order to investigate the possible existence
of a fixed point exhibiting this scale symmetry. A scale-invariant cutoff is
necessarily field-dependent. The simplified flow equation offers a tool for this
investigation.

Let us consider a flowing action of the form
\begin{equation}
\label{eq:KIN1}
\Gak=\int_x\sqrt{g}\left\{\frac12K(\vp)\partial^\mu\vp\partial_\mu\vp+U(\vp)\right\}\
.
\end{equation}
The second functional derivative reads
\begin{align}
\label{eq:KIN2}
g^{-1/2}\Gakt=&\,-\Big[K\partial^\mu\partial_\mu+\partial_\vp
K\partial^\mu\vp\partial_\mu+\frac12\partial_\vp^2K\partial^\mu\vp\partial_\mu\vp\nn\\
&\,+\partial_\vp K\partial^\mu\partial_\mu\vp\Big]+\partial_\vp^2U\ .
\end{align}
We define $\Gart$ by
\begin{equation}
\label{eq:KIN2A}
\Gakt=g^{1/2}K^{1/2}\Gart K^{1/2}\ ,
\end{equation}
where
\begin{equation}
\label{eq:KIN3}
\Gart=-\partial^\mu\partial_\mu-C+K^{-1}\partial_\vp^2U\ ,
\end{equation}
with
\begin{equation}
\label{eq:KIN4}
C=\frac14\gl\partial_\vp\ln
K\gr^2\partial^\mu\vp\partial_\mu\vp+\frac12\partial_\vp\ln
K\partial^\mu\partial_\mu\vp\ .
\end{equation}
One identifies in eq.~\eq{1B} $N=K^{1/2}$ and for eq.~\eq{7A}
$E=-\partial_t\ln K$. For a non-trivial field dependence of the kinetial the
field-dependent renormalization factor $N(\vp)$ in the simplified flow
equation~\eq{1A},~\eq{1B} plays an important role.

We next evaluate the flow equation for $K(\vp)$ from a derivative expansion of
the flowing action, using simple heat kernel methods. For the heat kernel
expansion we employ
\begin{align}
\label{eq:KIN5}
&\tr\Big\{\exp\big\{s\gl
g^{\mu\nu}\partial_\mu\partial_\nu-m^2(x)\gr\big\}\Big\}\nn\\
=&\,\frac1{16\pi^2}\int_x\sqrt{g}\exp\gl-sm^2(x)\gr\nn\\
&\quad\times\gl s^{-2}+c_1\partial^\mu\partial_\mu m^2+c_2s\partial^\mu
m^2\partial_\mu m^2+\dots\gr
\end{align}
with
\begin{equation}
\label{eq:KIN6}
c_1=-\frac16\ ,\quad c_2=\frac{1}{12}\ .
\end{equation}
The dots indicate terms with more than two derivatives or derivatives acting on
the metric. For $n=3$ eq.~\eq{5A} yields
\begin{align}
\label{eq:KIN7}
\dt\Gamma_k=&\,\frac{k^6}{16\pi^2}\left(1-\frac\eta2\right)\int_x\sqrt{g}\int_0^\infty\text{d}z\,z^{-3}\nn\\
\times\int\frac{\text{d}s}{2\pi i}&\,e^{s(z-k^2-m^2(x))}\gl
s^{-2}+c_1\partial^\mu\partial_\mu m^2+c_2s\partial^\mu m^2\partial_\mu
m^2\gr\nn\\
=&\,\frac{1-\eta/2}{16\pi^2}\int_x\sqrt{g}\bigg(\frac{k^6}{2(k^2+m^2)}+\frac{c_1\partial^\mu\partial_\mu
m^2k^6}{(k^2+m^2)^3}\nn\\
&\,+\frac{3c_2\partial^\mu m^2\partial_\mu m^2k^6}{(k^2+m^2)^4}\bigg)\ .
\end{align}
For homogeneous $m^2$ we recover the flow of the potential in eq.~\eq{A5}.

With $m^2=\partial_\vp^2U/K-C$ one employs for the second order of the
derivative expansion
\begin{equation}
\label{eq:KIN8}
\partial_\mu m^2=\gl\partial_\vp^3UK^{-1}-\partial_\vp^2U\partial_\vp
KK^{-2}\gr\partial_\mu\vp\ .
\end{equation}
Writing eq.~\eq{KIN7} in the form
\begin{equation}
\label{eq:35A}
\dt\Gamma_k=\int_x\sqrt{g}\Big\{\dt U+\frac12\dt
K\partial^\mu\vp\partial_\mu\vp\Big\}\ ,
\end{equation}
one obtains the flow of the potential
\begin{equation}
\label{eq:35B}
\dt U=\frac{(1-\eta/2)k^6}{32\pi^2(k^2+\partial_\vp^2UK^{-1})}\ .
\end{equation}
The flow equation for the kinetial is extracted by partial integration as
\begin{align}
\label{eq:KIN9}
\dt K=&\,\frac{(1-\eta/2)k^6}{16\pi^2}\bigg[\frac{(\partial_\vp\ln
K)^2-2\partial_\vp^2\ln K}{4(k^2+\partial_\vp^2UK^{-1})^2}\nn\\
&\,+\frac{(\partial_\vp^3UK^{-1}-\partial_\vp^2U\partial_\vp K
K^{-2})\partial_\vp\ln K}{(k^2+\partial_\vp^2UK^{-1})^3}\nn\\
&\,-\frac{(\partial_\vp^3UK^{-1}-\partial_\vp^2U\partial_\vp K
K^{-2})^2}{2(k^2+\partial_\vp^2UK^{-1})^4}\bigg]\ .
\end{align}
The simplified flow equation permits a straightforward computation of the flow
of both the potential and the kinetial for an arbitrary field-dependent kinetial
$K(\vp)$.

We are interested in possible scaling solutions of the
system~\eq{35B},~\eq{KIN9} which can be associated to fixed points. It is
striking that the flow equation~\eq{35A}-~\eq{KIN9} admits a scaling
solution with flat potential and kinetial $\sim\vp^{-2}$
\begin{equation}
\label{eq:KIN20}
U=u_0k^4\ ,\quad K=\frac{\kappa_0}{\vp^2}\ ,\quad u_0=\frac1{128\pi^2}\ .
\end{equation}
This fixed point realizes the exact global scale symmetry $\vp\to\beta\vp$.
Indeed, the simplified flow equation~\eq{1A} respects this symmetry. With
$\Gakt$ scaling $\sim\beta^{-2}$ this holds provided $N$ scales
$\sim\beta^{-1}$, such that $\Gart$ and $E$ are invariant. The field-dependence
of the cutoff $\cR_k\sim K\sim\vp^{-2}$ in eq.~\eq{MF1} is crucial for the
compatibility with this scale symmetry. This fixed point may be called ``kinetic
fixed point''. As a second fixed point one has the trivial fixed point for which
one has in eq.~\eq{KIN20} a constant kinetial $K=K_0$. This second fixed point
is invariant under a constant shift in $\vp$.

It will be interesting to find out if the kinetic fixed point with
multiplicative scale symmetry~\eq{KIN20} survives an extended truncation and
the inclusion of corrections to the simplified flow equation. If this is the
case, it offers the interesting prospect of a non-trivial UV-complete scalar
field theory in four dimensions. One expects the existence of crossover
trajectories between the scale invariant and the trivial fixed point. For models
on the crossover trajectory one expects the presence of interactions. For a
crossover trajectory any point on the trajectory can be realized as a
UV-complete theory. If a fixed point with the properties~\eq{KIN20} can be
established, this may overcome the old ``triviality problem'' for scalar
theories in four dimensions.

\subsection*{Exact flow equation for field-dependent cutoff}

For a cutoff function $R_k$ which depends on the macroscopic field $\vp$ the
flow equation~\eq{MF1} is no longer exact. There are correction terms
$~\sim\partial_\vp R_k$ that we will derive next. In a loop expansion they are
of higher order. An optimization of the definition of the flowing action
minimizes the importance of the correction term. For this purpose we define the
flowing action $\Gak[\vp]$ by a functional identity
\begin{equation}
\label{eq:EF1}
\Gak[\vp]=-\ln\int\cD\chi\,\exp\big\{-\bar S_k[\chi,\vp]\big\}\ ,
\end{equation}
where $\bar S[\chi,\vp]$ is related to the classical action $S[\chi]$ by the
addition of a cutoff and a source term,
\begin{align}
\label{eq:EF2}
\bar S_k[\chi,\vp]=&\,S[\chi]+\frac12(\chi-\vp)R_k(\vp)(\chi-\vp)\nn\\
&-\left(\frac{\partial\Gamma_k}{\partial\vp}+L_k(\vp)\right)(\chi-\vp)\ .
\end{align}
We employ here a notation where $\vp$ stands for a vector with components
$\vp_\alpha$, with $\alpha$ labeling both position or momentum and internal
indices. The cutoff $R_k$ is a matrix with elements $(R_k)_{\alpha\beta}$, and
products of vectors stand for scalar products, $\chi\vp=\chi_\alpha\vp_\alpha$.
For $k\to0$ both $R_k$ and $L_k$ vanish and $\Gamma_{k\to0}$ equals the quantum
effective action, as usually defined by a Legendre transform of the Schwinger
functional. For $\partial R_k/\partial\vp=0$ one has $L_k=0$ and our definition
of the flowing action equals the effective average action~\cite{CWEFE}.

We define $L_k(\vp)$ such that the macroscopic field $\vp$ equals the
expectation value of the microscopic field
\begin{equation}
\label{eq:EF3}
\langle\chi\rangle=\vp\ .
\end{equation}
Here expectation values are formed with the weight factor $\exp\gl-\bar S_k\gr$,
\begin{equation}
\label{eq:EF4}
\langle A\rangle=\frac{\int\cD\chi\,Ae^{-\bar S_k}}{\int\cD\,e^{-\bar S_k}}\ .
\end{equation}
The logarithmic $k$-derivative of eq.~\eq{EF1} at fixed $\vp$ yields the
exact flow equation
\begin{equation}
\label{eq:EF5}
\partial_t\Gak=\frac12\tr\big\{\partial_tR_kG\big\}\ ,
\end{equation}
which involves the connected two-point function
\begin{equation}
\label{eq:EF6}
G_{\alpha\beta}=\langle(\chi_\alpha-\vp_\alpha)(\chi_\beta-\vp_\beta)\rangle\ .
\end{equation}

We need an expression for $L_k(\vp)$ and a relation between $G$ and functional
derivatives of $\Gamma_k$. Taking a derivative of eq.~\eq{EF1} with respect
to $\vp_\alpha$ yields
\begin{align}
\label{eq:EF7}
\frac{\partial\Gak}{\partial\vp_\alpha}=&\,\left\langle\frac{\partial\bar
S_k}{\partial\vp_\alpha}\right\rangle=\frac{\partial\Gak}{\partial\vp_\alpha}-M_{\alpha\beta}(\langle\chi_\beta\rangle-\vp_\beta)\nn\\
&+L_{k,\alpha}+\frac12\tr\left\{\frac{\partial
R_k}{\partial\vp_\alpha}G\right\}\ ,
\end{align}
with
\begin{equation}
\label{eq:EF8}
M_{\alpha\beta}=\gl\Gakt+R_k\gr_{\alpha\beta}+\frac{\partial
L_{k,\beta}}{\partial\vp_\alpha}\ .
\end{equation}
The relation~\eq{EF3} is obeyed for the choice
\begin{equation}
\label{eq:EF9}
L_{k,\alpha}=-\frac12\tr\left\{\frac{\partial R_k}{\partial\vp_\alpha}G\right\}\
.
\end{equation}

Taking further a $\vp$-derivative of eq.~\eq{EF3} one finds
\begin{align}
\label{eq:EF10}
\delta_{\alpha\beta}=&\,\frac{\partial\langle\chi_\beta\rangle}{\partial\vp_\alpha}=-\left\langle\frac{\partial\bar
S_k}{\partial\vp_\alpha}(\chi_\beta-\vp_\beta)\right\rangle\nn\\
=&\,M_{\alpha\gamma}G_{\gamma\beta}-\frac12\frac{\partial
R_{k,\gamma\delta}}{\partial\vp_\alpha}H_{\gamma\delta\beta}\ ,
\end{align}
which involves the connected three-point function
\begin{equation}
\label{eq:EF11}
H_{\gamma\delta\beta}=\big\langle(\chi_\gamma-\vp_\gamma)(\chi_\delta-\vp_\delta)(\chi_\beta-\vp_\beta)\big\rangle\
.
\end{equation}
One infers the identity
\begin{equation}
\label{eq:EF12}
G=\gl\Gakt+R_k+Q_k\gr^{-1}\ ,
\end{equation}
with
\begin{equation}
\label{eq:EF13}
Q_{k,\alpha\beta}=-\frac12\left[\frac{\partial}{\partial\vp_\alpha}\tr\left\{\frac{\partial
R_k}{\partial\vp_\beta}G\right\}+\tr\left\{\frac{\partial
R_k}{\partial\vp_\alpha}\tilde H_\beta\right\}\right]\ .
\end{equation}
Here the matrix $\tilde H_\beta$ has elements
\begin{equation}
\label{eq:EF14}
\tilde H_{\beta,
\gamma\delta}=H_{\gamma\delta\varepsilon}G^{-1}_{\varepsilon\beta}\ .
\end{equation}
Since the matrices $G$, $\Gakt$ and $R_k$ are symmetric, we symmetrize $Q_k$.

We arrive at the final form of the exact flow equation
\begin{equation}
\label{eq:EF15}
\partial_t\Gak=\frac12\tr\Big\{\dt R_k\gl\Gakt+R_k+Q_k\gr^{-1}\Big\}\ .
\end{equation}
The corrections to the flow equation~\eq{MF1} appear in the form of a
field-dependent correction $Q_k$ to the infrared cutoff $R_k$ in the effective
$k$- and field-dependent propagator. As long as $Q_k$ remains small as compared
to $\Gakt+R_k$ one only finds small corrections. In order to establish the
reliability of the simplified flow equation we have to compute the size of
$Q_k$. This is done in the following. While somewhat technical, we consider this
part as an important progress as compared to earlier papers on the gauge
invariant flow equation where only qualitative estimates have been given.

\subsection*{Corrections to the simplified flow equation}

In order to close the exact flow equation~\eq{EF15} we need to express $Q_k$
in terms of $\Gak$ and $R_k$. This closure needs a relation between $\tilde
H_\beta$ and $G$ or $Q$ and an expression for $\partial\tilde
G/\partial\vp_\alpha$. One may take a $\vp$-derivative of the relation,
\begin{equation}
\label{eq:EF16}
G_{\gamma\delta}\gl\Gakt+R_k+Q_k\gr_{\delta\varepsilon}=\delta_{\gamma\varepsilon}\
,
\end{equation}
which expresses $\partial_\alpha G_{\gamma\delta}$ in terms of the third
functional derivative of $\Gak$,
\begin{align}
\label{eq:EF17}
\partial_\alpha
G_{\gamma\delta}=&\,-G_{\gamma\varepsilon}\partial_\alpha\gl\Gakt+R_k+Q_k\gr_{\varepsilon\eta}G_{\eta\delta}\
,\nn\\
\partial_\alpha\Gamma^{(2)}_{k,\varepsilon\eta}=&\,\partial_\alpha\partial_\varepsilon\partial_\eta\Gak=\gl\Gamma^{(2)}_k\gr_{\alpha\varepsilon\eta}\
.
\end{align}
(We use here and from now on a notation $\partial_\alpha F=\partial
F/\partial\vp_\alpha$.) On the other hand, we may take a $\vp$-derivative of the
defining relation~\eq{EF6},
\begin{align}
\label{eq:EF14}
\partial_\alpha G_{\gamma\delta}=&\,-\big\langle\partial_\alpha\bar
S_k\big[(\chi_\gamma-\vp_\gamma)(\chi_\delta-\vp_\delta)-
G_{\gamma\delta}\big]\big\rangle\\
=&\,M_{\alpha\eta}H_{\eta\gamma\delta}-\frac12\partial_\alpha
R_{k,\varepsilon\eta}\gl
V_{\varepsilon\eta\gamma\delta}-G_{\varepsilon\eta}G_{\gamma\delta}\gr\nn\ ,
\end{align}
which contains the four-point function
\begin{equation}
\label{eq:EF15B}
V_{\varepsilon\eta\gamma\delta}=\big\langle(\chi_\varepsilon-\vp_\varepsilon)(\chi_\eta-\vp_\eta)(\chi_\gamma-\vp_\gamma)(\chi_\delta-\vp_\delta)\big\rangle\
.
\end{equation}

Writing $\partial_\alpha G_{\gamma\delta}$ in the form
\begin{align}
\label{eq:EF16}
\partial_\alpha G_{\gamma\delta}=&\,\gl\tilde
H_\alpha\gr_{\gamma\delta}+N_{\alpha\gamma\delta}\ ,\nn\\
N_{\alpha\gamma\delta}=&\,\frac12\tr\big\{\partial_\alpha R_k\tilde
H_\eta\big\}H_{\eta\gamma\delta}-\frac12\tr\big\{\partial_\alpha R_k\tilde
V_{\gamma\delta}\big\}\nn\\
&+\frac12\tr\big\{\partial_\alpha R_kG\big\}G_{\gamma\delta}\ ,\nn\\
\gl\tilde
V_{\gamma\delta}\gr_{\varepsilon\eta}=&\,V_{\varepsilon\eta\gamma\delta}\ ,
\end{align}
one infers for $Q_k$ the expression
\begin{align}
\label{eq:EF17}
\gl Q_k\gr_{\alpha\beta}=&\,-\frac12\tr\big\{\partial_\alpha\partial_\beta
R_kG+\partial_\alpha R_k\partial_\beta G+\partial_\beta R_k\partial_\alpha
G\big\}\nn\\
&+D_{k,\alpha\beta}\ ,
\end{align}
with
\begin{equation}
\label{eq:EF18}
D_{k,\alpha\beta}=\frac12\partial_\alpha
R_{k,\gamma\delta}N_{\beta\gamma\delta}\ .
\end{equation}
This procedure does not close since $D_k$ involves the unknown four-point
function $V$ through eq.~\eq{EF16}. One could continue by taking higher
$\vp$-derivatives.

We may truncate $D_k=0$ and consider the resulting form of $Q_k$ in
eq.~\eq{EF17} as a first term in some expansion. Higher orders in this
expansion take higher $\vp$-derivatives and include the effects of higher
correlation functions. This expansion leads to a type of loop expansion for
$Q_k$. Each trace amounts to a momentum integral in a loop. We conclude that
$Q_k$ is of one-loop order, while $D_k$ involves two loops since
$N_{\beta\gamma\delta}$ is already of one-loop order. Each loop order implies an
additional factor $\partial_\alpha R_k$. We observe that $Q_k$ is
loop-suppressed as compared to $\Gakt+R_k$.

In lowest order of an iterative computation of $Q_k$ we can neglect $D_k$ and
approximate in eq.~\eq{EF17} $G$ by $\gl\Gakt+R_k\gr^{-1}$,
\begin{align}
\label{eq:EF19}
Q_k=&\,Q_k^{(1)}+Q_k^{(2)}\ ,\nn\\
Q^{(1)}_{k,\alpha\beta}=&\,-\frac12\tr\Big\{\partial_\alpha\partial_\beta
R_k\gl\Gakt+R_k\gr^{-1}\Big\}\ ,\\
Q^{(2)}_{k,\alpha\beta}=&\,-\frac12\tr\Big\{\partial_\alpha
R_k\partial_\beta\gl\Gakt+R_k\gr^{-1}+\alpha\leftrightarrow\beta\Big\}\ .\nn
\end{align}
The correction $Q^{(2)}_k$ has a term proportional to the third derivative of
$\Gamma_k$,
$\partial_\beta\gl\Gakt\gr_{\gamma\delta}=\Gamma^{(3)}_{k,\beta\gamma\delta}$.

The correction $Q_k$ in the exact flow equation~\eq{EF15} fits well into the
structure of the simplified flow equation. One simply replaces $\Gakt$ by
$\Gakt+Q_k$. Correspondingly, we choose the cutoff function such that $r$ is now
a function of $x$
\begin{equation}
\label{eq:EF20}
x=k^{-2}g^{-1/2}N^{-1}\gl\Gakt+Q_k\gr\gl N^T\gr^{-1}\ .
\end{equation}
This replaces in the flow equation~\eq{1A} the matrix $\Gart$ by
$k^2x=\Gart+g^{-1/2}N^{-1}Q_k\gl N^T\gr^{-1}$. Including the correction the flow
equation~\eq{1A} is extended to
\begin{equation}
\label{eq:EF21}
\dt\Gak=\tr\Big\{\gl1-\frac12E\gr\gl B+x_Q+1\gr^{-n}\Big\}\ ,
\end{equation}
where
\begin{equation}
\label{eq:EF22}
x_Q=g^{-1/2}k^{-2}N^{-1}Q_k\gl N^T\gr^{-1}\ .
\end{equation}
As long as $x_Q$ remains small as compared to $1+B$ the leading order
approximation~\eq{1A} is valid. In the limit of constant $x_Q$ the analytic
structure of the simplified flow equation only changes by a shift of the poles
to $B=-(1+x_Q)$. This may be seen as an ``additive renormalization'' of $k^2$ in
the regularized propagator.

\subsection*{Lowest order correction}

In lowest order the correction $x_Q$ inserts in eq.~\eq{EF22} the
approximation~\eq{EF19}. For an estimate of the size of $x_Q$ we neglect
first $\partial_\alpha N$, $\partial_\alpha g$ as well as commutators of $x$ and
$\partial_\alpha x$, such that
\begin{equation}
\label{eq:EF23}
\partial_\alpha\partial_\beta R_k=g^{1/2}k^2N\gl\partial_\alpha\partial_\beta
xr'+\partial_\alpha x\partial_\beta xr''\gr N^T\ ,
\end{equation}
where primes denote here derivatives with respect to $x$. In leading order we
can replace $\partial_\beta x$ by $\partial_\beta B$ and obtain
\begin{equation}
\label{eq:EF24}
Q^{(1)}_{k,\alpha\beta}=-\frac12\tr\Big\{\gl\partial_\alpha\partial_\beta
Br'+\partial_\alpha B\partial_\beta Br''\gr\gl B+r\gr^{-1}\Big\}\ .
\end{equation}
The expression~\eq{EF24} can be related to the renormalized dimensionless
1PI-three- and four-point functions
\begin{align}
\label{eq:EF25}
\Gamma^{(3)}_{R,\alpha\gamma\delta}=&\,k^2N^{-1}_{\alpha\alpha'}\partial_{\alpha'}
B_{\gamma\delta}\ ,\nn\\
\Gamma^{(4)}_{R,\alpha\beta\gamma\delta}=&\,k^2N^{-1}_{\alpha\alpha'}N^{-1}_{\beta\beta'}\partial_{\alpha'}\partial_{\beta'}B_{\gamma\delta}\
,
\end{align}
implying
\begin{align}
\label{eq:EF26}
x^{(1)}_{Q,\alpha\beta}=&\,-\frac12g^{-1/2}\bigg\{k^{-4}\Gamma^{(4)}_{R,\alpha\beta\gamma\delta}\big[r'(B+r)^{-1}\big]_{\delta\gamma}\nn\\
&+k^{-6}\Gamma^{(3)}_{R,\alpha\gamma\varepsilon}\Gamma^{(3)}_{R,\beta\varepsilon\delta}\big[r''(B+r)^{-1}\big]_{\delta\gamma}\bigg\}\
.
\end{align}
The momentum integrals of the loop corresponding to the trace (sum over $\gamma$
,$\delta$) are both ultraviolet and infrared finite since $r'$ or $r''$ decay
rapidly for large $x$, and $r$ approaches a constant for $x\to0$. For
homogeneous macroscopic fields $x_Q$ is diagonal in momentum space, similar to
$r$. As compared to $r$ it is suppressed by a loop factor $(16\pi^2)^{-1}$ and
it involves the renormalized couplings~\eq{EF25}. We conclude that unless the
renormalized couplings~\eq{EF25} are very large the corrections to the
simplified flow equation~\eq{1A} are small and the iterative loop expansion
seems to converge. The situation for the contribution from $Q_k^{(2)}$ is
similar.

As an example we consider a single scalar field. For a homogeneous macroscopic
field $\vp$ one has in momentum space
\begin{align}
\label{eq:EF27}
\Gamma_R^{(4)}(p_1,p_2,p_3,p_4)=&\,3\lambda_R\delta(p_1-p_2+p_3-p_4)\ ,\nn\\
\Gamma_R^{(2)}(p_1,p_2)=&\,\gl p_1^2+m_R^2\gr\delta(p_1-p_2)\ .
\end{align}
Both $m_R^2$ and $\lambda_R$ may depend on $\vp$, while we assume here for
simplicity that they do not depend on momentum. We define
\begin{align}
\label{eq:EF28}
&\big[r'(B)\gl B+r(B)\gr^{-1}\big](p_1,p_2)\nn\\
=&\,f(B)(p_1,p_2)=f\left(\frac{p_1^2+m_R^2}{k^2}\right)\delta(p_1-p_2)\ ,
\end{align}
and obtain for the first term in eq.~\eq{EF26}
\begin{align}
\label{eq:EF29}
\tilde
x^{(1)}_Q(p_3,p_4)=&\,-\frac32g^{-1/2}k^{-4}\lambda_R\delta(p_3-p_4)\int_qf\left(\frac{q^2+m_R^2}{k^2}\right)\nn\\
=&\,-\frac{3\lambda_R}{16\pi^2}h\left(\frac{m_R^2}{k^2}\right)\delta(p_3-p_4)\ .
\end{align}
With $\tilde m^2=m_R^2/k^2$ one has
\begin{equation}
\label{eq:EF30}
h(\tilde m^2)=\int_{\tilde m^2}^\infty\text{d}x'\,(x'-\tilde
m^2)\partial_{x}r(x')\gl x'+r(x')\gr^{-1}\ .
\end{equation}
The function $h(\tilde m^2)$ becomes a constant for $\tilde m^2=0$, and
decreases rapidly for $\tilde m^2\gg1$ due to the fast decay of $\partial_x r$
in eq.~\eq{N3}. We observe the expected suppression factor
$\lambda_R/(16\pi^2)$. As compared to $B+1\approx q^2/k^2+\tilde m^2+1$ the
correction $\tilde x^{(1)}_Q$ becomes tiny for large $\tilde m^2$ and affects
the momentum integral~\eq{EF21} only for $q^2\approx k^2$. Similar
considerations hold for the other parts of the correction term.

We conclude that the correction term has only a small influence on the flow
equation for the effective potential as long as $\lambda_R/(16\pi^2)$ remains
small. For quantitative precision it cannot be neglected, however. Besides the
small factor $\lambda_R/(16\pi^2)$ the relative contribution to the
$\vp$-dependence of the potential is suppressed by $\partial h/\partial\mtil^2$.

\subsection*{Robustness of kinetic fixed point}

Let us next ask how corrections affect the possible kinetic fixed
point~\eq{KIN20}. If the classical action is invariant under the constant
scaling $\chi\to\beta\chi$, we can choose a cutoff $R_k$ which scales
$\sim\beta^{-2}$, e.g. $\sim\vp^{-2}$. If the flowing action $\Gamma_k$ is
invariant under this rescaling up to a field-independent constant, one has
$\partial\Gamma_k/\partial\vp\sim\beta^{-1}$ and $L_k\sim\beta^{-1}$. One
concludes that $\Sbar_k$ in eq.~\eq{EF2} is scale invariant, and therefore
$\Gamma_k$ is scale invariant up to a field- and $k$-independent constant which
arises from the measure $\int\cD\chi$ in eq.~\eq{EF1}. Thus the exact flow
equation~\eq{EF5},~\eq{EF15} is scale invariant. One may verify that the
correction term does not perturb scale invariance since $Q_k$ scales
$\sim\beta^{-2}$.

From the scale invariance of the exact flow equation for scale invariant
$\Gamma_k$ one infers that the manifold of scale-invariant flowing actions
constitutes a partial fixed point (or better fixed surface). Starting at a point
on this surface all flow trajectories remain within this surface. For the
investigation of a scale-invariant fixed point a restriction to this surface is
therefore sufficient. This implies immediately a flat potential and a kinetial
$K=\kappa(k)\vp^{-2}$. An exact fixed point requires that the flow of
$\kappa(k)$ has a fixed point.

With respect to the global scaling $\vp\to\beta\vp$ with invariant $g_{\mu\nu}$
an invariant flowing action takes in flat space the general form
\begin{equation}
\label{eq:KK1}
\Gamma_k=\int_x\sqrt{g}\left\{\frac{\kappa(k)}{2\vp^2}\partial^\mu\vp\partial_\mu\vp+U_0(k)+\dots\right\}\
,
\end{equation}
where dots denote terms with four or more derivatives of $\ln(\vp)$. With
$N[\vp]=\sqrt{\kappa}/|\vp|$, $E[\vp]=-\dt\ln\kappa=\eta$, one finds that
$\Gart$ is independent of $\kappa$ and $k$,
\begin{equation}
\label{eq:KK2}
\Gart=-\partial^\mu\partial_\mu+\partial^\mu\partial_\mu\ln\vp\ ,
\end{equation}
such that eq.~\eq{MF3} holds. In lowest order the simplified flow equation
yields with with eq.~\eq{KIN9} a vanishing flow $\partial_t\kappa=0$,
implying a line of fixed points~\eq{KIN20} with free parameter $\kappa_0$.

The correction term adds to $\Gart$ in eq.~\eq{KIN3} a contribution
\begin{align}
\label{eq:KK3}
\tilde Q_k=&\,g^{-1/2}\kappa^{-1}\vp Q_k\vp\nn\\
=&\,\frac{k^2}{16\pi^2\kappa}\big(f_1k^2+f_2\partial^\mu\ln\vp\partial_\mu+f_3\partial^\mu\ln\vp\partial_\mu\ln\vp\nn\\
&\quad\quad\quad+f_4\partial^\mu\partial_\mu\ln\vp\big)\ ,
\end{align}
with field-independent dimensionless functions
$f_j(-\partial^\mu\partial_\mu/k^2)$, and $x_Q=\tilde Q/k^2$. Here the general
form of the second expression uses scale invariance of $\tilde Q_k$ and
truncates in second order of a derivative expansion. As an example, we take
$f_2=0$ and constant $f_1$, $f_3$, $f_4$. Then the terms in eq.~\eq{KIN7}
which involve $\partial_\mu\tilde Q_k$ contain more than two derivatives and do
not contribute to the flow of $K$. The kinetial flows only for $f_3\neq0$,
\begin{equation}
\label{eq:KK4}
\partial_t\kappa=-\frac{f_3k^4}{(16\pi^2)^2\kappa}\left(1-\frac\eta2\right)
\left(1+\frac{f_1k^2}{16\pi^2\kappa}\right)^{-2}\ .
\end{equation}

Counting factors of $\kappa$ one finds that $Q_k$ has no factor $\kappa$ and
therefore the functions $f_j$ are independent of $\kappa$. The coefficients
$f_j$ can be computed in lowest order from eq.~\eq{EF19}. With
$N=\sqrt{\kappa}/\vp$ the terms $\sim\partial_\alpha N$ can no longer be
neglected. Independently of the precise values of $f_j$ the flow of $\kappa$
becomes very slow for large $\kappa/k^2$. There is an exact fixed point at
$\kappa^{-1}=0$. The line of fixed points with arbitrary $\kappa_0$ found in
lowest order remains valid for $f_3=0$, or approximately valid for large
$\kappa/k^2$. Once in the range of small $k^2/\kappa$ the flow of $\kappa$ stops
as $k$ decreases towards zero, $\kappa\approx\kappa_0+ck^4/\kappa_0$. The flow
equation for $v=k^2/\kappa$ has an infrared attractive fixed point for $v=0$. It
could have an additional infrared fixed point for finite $v$, for example if
$f_3$ is negative. In any case, if one starts in the range of attraction of the
fixed point at $v=0$ the flow will always reach small values of $v$.

We conclude that the kinetic fixed point~\eq{KIN20} is either exact with
arbitrary $\kappa_0$, or becomes a very good approximation for small enough
$k^2$. For establishing this result it is crucial that the simplified flow
equation and its corrections respect the symmetry of multiplicative scaling of
the scalar field. This constitutes an important strength of our approach, which
is not easy to realize for other forms of flow equations.

\subsection*{Local gauge symmetry}

The simplified flow equation can be seen as a particular version of the gauge
invariant flow equation~\cite{CWGIFE, CWGIYM}. The central ingredient is the
field-dependence of the cutoff function $R_k(\vp)$ in the functional
integral~\eq{EF1},~\eq{EF2}. This permits us to choose $R_k$ as a function
of the covariant Laplacian or other operators that transform homogeneously as
tensors with respect to local gauge transformations. In particular, for a gauge
invariant flowing action the second functional derivative $\Gakt$ transforms as
a tensor. In consequence, $R_k$ should transform as a tensor as well. This
requires that $R_k$ has to depend on the macroscopic gauge field appearing in
covariant Laplacians. The simplified flow equation obeys this requirement for
gauge invariance.

We next establish the gauge invariance of the flowing action and the
corresponding exact flow equation~\eq{EF15} by using the
definition~\eq{EF1},~\eq{EF2}. The difference
between the microscopic and the macroscopic gauge field $\chi_\alpha-\vp_\alpha$
transforms homogeneously as a vector, and the correlation function
$G_{\alpha\beta}$ is therefore a tensor. One concludes that $L_{k,\alpha}$ in
eq.~\eq{EF9} transforms as a vector, similar to
$\partial\Gamma_k/\partial\vp_\alpha$. For a gauge invariant classical action
$S[\chi]$ all pieces on the r.h.s. of eq.~\eq{EF2} are gauge invariant,
implying in turn the gauge invariance of the flowing effective
action~\eq{EF1} and the exact flow equation~\eq{EF5}. The exact correction
$Q_k$ in eq.~\eq{EF13} transforms as a tensor, implying gauge invariance of
the exact flow equation~\eq{EF15}. For the expansion of $Q_k$ the lowest
order~\eq{EF19} transforms as a tensor, leading for the flow equation to
gauge invariance of each order in the expansion.

For a continuum formulation the functional integral for local gauge symmetries
needs a regularization, typically implemented by a gauge fixing term with
associated Faddeev-Popov determinant. We focus on the physical gauge fixing (for
Yang-Mills theories this is Landau gauge) which realizes effectively a
projection in the physical fluctuation modes~\cite{CWGIFE, CWGIYM}. Implementing
a cutoff for the gauge-fluctuations and the Faddeev-Popov determinant adds to
the flow
equations~\eq{1A},~\eq{MF1},~\eq{EF5},~\eq{EF15},~\eq{EF21} a
universal ``measure term''. This measure term does not depend on $\Gamma_k$,
reflecting that it results ultimately from the regularization of the measure in
the functional integral. It is a function of covariant derivatives. Its explicit
form can be found in ref.~\cite{CWGIFE, CWGIYM}. For the case of quantum gravity
the coincidence of the measure term with a corresponding term in the background
formalism with physical gauge fixing has been discussed explicitly in
ref.~\cite{PRWY}. The main term in the flow equation besides the measure term
involves effectively only the physical fluctuations (e.g. transverse gauge
fields for Yang-Mills theories). Correspondingly, the cutoff function involves
the projection of the covariant Laplacian on the physical modes, or
generalizations thereof as $\Gakt$.

We conclude that gauge invariance of the flow equation holds whenever $R_k$
transforms homogeneously as a tensor. What is new in the present note is the
explicit computation of the corrections to the flow equation~\eq{MF1}. This
correction is of higher loop order. For the case of Yang-Mills theories with a
small renormalized gauge coupling $g_R$ one expects that the correction is
suppressed as compared to the leading term by a factor of $g_R^2/(16\pi^2)$. The
additional factor of $g_R^2$ in the beta-function for the running gauge coupling
can be seen rather easily. In leading order one has $\Gakt=Z_kD$, with $D$ a
(modified) covariant Laplacian and $Z_k=g_R^{-2}$ the wave function
renormalization multiplying in $\Gamma_k$ the gauge invariant kinetic term
$F_{\mu\nu}F^{\mu\nu}$. With $R_k\sim Z_k$ the leading term in the flow
equation~\eq{MF1} does not involve a factor $Z_k$ on the r.h.s., leading to
\begin{equation}
\label{eq:GG1}
\dt Z_k=\frac c{16\pi^2}\ ,\quad \dt g_R^2=-\frac c{16\pi^2}g_R^4\ .
\end{equation}
The correction term $Q_k$ in eq~\eq{EF13},~\eq{EF19} contains no factor
$Z_k$. We may write is as $Q_k=\big[Z_kg_R^2/(16\pi^2)\big]q_k$ which makes the
suppression factor $g_R^2/(16\pi^2)$ as compared to $\Gakt+R_k$ directly
visible. The correction factor contributes to $\dt g_R^2$ only in the order
$g_R^6$.

We finally turn to the specific choice of $R_k$ leading to the simplified flow
equation~\eq{1A} or its correction~\eq{EF21}. If $N(\vp)$ is chosen to
transform homogeneously as a tensor also $\Gart$ and $E[\vp]$ transform as
tensors, implying the gauge invariance of the flow equations. In leading order
for a Yang-Mills theory one has $N=Z_k^{1/2}$ and therefore
$E=-cg_R^2/(16\pi^2)$. The simple factor $(1-E/2)$ in eq.~\eq{1A} accounts
for more than $90\%$ of the two-loop contribution to $\dt g_R^2$~\cite{RWGT}.
The combined effects of extended truncations and corrections to eq.~\eq{1A}
are only a rather small effect for small gauge couplings.

\subsection*{Discussion}

In this note we propose a version of a simplified functional flow equation which
is gauge invariant, admits analytic continuation between euclidean and Minkowski
signature, and guarantees that the cutoff removes the critical fluctuations. The
first ingredient for this simplified flow equation is the flow
equation~\eq{MF1} for a field-dependent cutoff. At this level consistency
with all symmetries, in particular local gauge symmetries, can be achieved. We
consider eq.~\eq{MF1} here as an approximation, to which we have computed
corrections in the form of a loop expansion. This loop expansion should not be
confounded with a perturbative expansion of the flow generator in a small
coupling. Already the lowest order~\eq{MF1} constitutes a highly
non-perturbative setting. The loop expansion only concerns the computations of
corrections to the effective cutoff in case where the latter is taken to be
field-dependent. It has been argued that eq.~\eq{MF1} may become exact for a
further optimization of the flowing effective action~\cite{CWGIFE}. This
requires the existence of solutions of rather complex functional differential
equations. In view of its complexity we leave this issue open and stick here to
the definition~\eq{EF1}~\eq{EF2} and associated exact flow
equation~\eq{EF15}.

The second ingredient for the simplified flow equation concerns a particular
choice of the cutoff function. This leads to the simple analytic structure and
to an effective cutoff for critical fluctuations in a rather general context.
This choice of the cutoff function is only possible for a field-dependent
cutoff.

Our computation of the form of the correction terms to the flow
equation~\eq{MF1} yields further support for the (approximate) validity of
computations with gauge invariant flow equations. These flow equations have been
used for a rather simple understanding of flowing coupling functions in quantum
gravity~\cite{CWYA, PRWY, ESPA, CWQGS}. We have employed here the simplified
flow equation for the computation of the flow of the kinetial in
four-dimensional scalar field theories. This has revealed the possible existence
of a new kinetic fixed point which realizes the quantum scale symmetry with
respect to a multiplicative rescaling of the scalar field at fixed metric and
coordinates. A crossover from this fixed point to the trivial fixed point could
overcome the triviality problem for scalar quantum field theories in four
dimensions.

In our view the main strength of the proposed flow equation resides in its
qualitative robustness close to physical intuition, while maintaining a simple
form. This makes it suitable for many investigations of qualitative features of
non-perturbative systems, while high quantitative accuracy may not be the main
goal.


\subsection*{Acknowledgment:}
The author would like to thank J. Pawlowski for useful comments.




\nocite{*}
\bibliography{refs}

\begin{thebibliography}{49}%
\makeatletter
\providecommand \@ifxundefined [1]{%
 \@ifx{#1\undefined}
}%
\providecommand \@ifnum [1]{%
 \ifnum #1\expandafter \@firstoftwo
 \else \expandafter \@secondoftwo
 \fi
}%
\providecommand \@ifx [1]{%
 \ifx #1\expandafter \@firstoftwo
 \else \expandafter \@secondoftwo
 \fi
}%
\providecommand \natexlab [1]{#1}%
\providecommand \enquote  [1]{``#1''}%
\providecommand \bibnamefont  [1]{#1}%
\providecommand \bibfnamefont [1]{#1}%
\providecommand \citenamefont [1]{#1}%
\providecommand \href@noop [0]{\@secondoftwo}%
\providecommand \href [0]{\begingroup \@sanitize@url \@href}%
\providecommand \@href[1]{\@@startlink{#1}\@@href}%
\providecommand \@@href[1]{\endgroup#1\@@endlink}%
\providecommand \@sanitize@url [0]{\catcode `\\12\catcode `\$12\catcode
  `\&12\catcode `\#12\catcode `\^12\catcode `\_12\catcode `\%12\relax}%
\providecommand \@@startlink[1]{}%
\providecommand \@@endlink[0]{}%
\providecommand \url  [0]{\begingroup\@sanitize@url \@url }%
\providecommand \@url [1]{\endgroup\@href {#1}{\urlprefix }}%
\providecommand \urlprefix  [0]{URL }%
\providecommand \Eprint [0]{\href }%
\providecommand \doibase [0]{https://doi.org/}%
\providecommand \selectlanguage [0]{\@gobble}%
\providecommand \bibinfo  [0]{\@secondoftwo}%
\providecommand \bibfield  [0]{\@secondoftwo}%
\providecommand \translation [1]{[#1]}%
\providecommand \BibitemOpen [0]{}%
\providecommand \bibitemStop [0]{}%
\providecommand \bibitemNoStop [0]{.\EOS\space}%
\providecommand \EOS [0]{\spacefactor3000\relax}%
\providecommand \BibitemShut  [1]{\csname bibitem#1\endcsname}%
\let\auto@bib@innerbib\@empty
\bibitem [{\citenamefont {Wilson}(1971)}]{WIL}%
  \BibitemOpen
  \bibfield  {author} {\bibinfo {author} {\bibfnamefont {K.~G.}\ \bibnamefont
  {Wilson}},\ }\bibfield  {title} {\bibinfo {title} {Renormalization group and
  critical phenomena. i. renormalization group and the kadanoff scaling
  picture},\ }\href@noop {} {\bibfield  {journal} {\bibinfo  {journal} {Phys.
  Rev. B}\ }\textbf {\bibinfo {volume} {4}},\ \bibinfo {pages} {3174} (\bibinfo
  {year} {1971})}\BibitemShut {NoStop}%
\bibitem [{\citenamefont {Wegner}\ and\ \citenamefont {Houghton}(1973)}]{WEG}%
  \BibitemOpen
  \bibfield  {author} {\bibinfo {author} {\bibfnamefont {F.~J.}\ \bibnamefont
  {Wegner}}\ and\ \bibinfo {author} {\bibfnamefont {A.}~\bibnamefont
  {Houghton}},\ }\bibfield  {title} {\bibinfo {title} {Renormalization group
  equation for critical phenomena},\ }\href@noop {} {\bibfield  {journal}
  {\bibinfo  {journal} {Phys. Rev. A}\ }\textbf {\bibinfo {volume} {8}},\
  \bibinfo {pages} {401} (\bibinfo {year} {1973})}\BibitemShut {NoStop}%
\bibitem [{\citenamefont {Wilson}\ and\ \citenamefont {Kogut}(1974)}]{WILKO}%
  \BibitemOpen
  \bibfield  {author} {\bibinfo {author} {\bibfnamefont {K.~G.}\ \bibnamefont
  {Wilson}}\ and\ \bibinfo {author} {\bibfnamefont {J.}~\bibnamefont {Kogut}},\
  }\bibfield  {title} {\bibinfo {title} {The renormalization group and the
  epsilon expansion},\ }\href@noop {} {\bibfield  {journal} {\bibinfo
  {journal} {Physics Reports}\ }\textbf {\bibinfo {volume} {12}},\ \bibinfo
  {pages} {75} (\bibinfo {year} {1974})}\BibitemShut {NoStop}%
\bibitem [{\citenamefont {Polchinski}(1984)}]{POL}%
  \BibitemOpen
  \bibfield  {author} {\bibinfo {author} {\bibfnamefont {J.}~\bibnamefont
  {Polchinski}},\ }\bibfield  {title} {\bibinfo {title} {Renormalization and
  effective lagrangians},\ }\href@noop {} {\bibfield  {journal} {\bibinfo
  {journal} {Nuclear Physics B}\ }\textbf {\bibinfo {volume} {231}},\ \bibinfo
  {pages} {269} (\bibinfo {year} {1984})}\BibitemShut {NoStop}%
\bibitem [{\citenamefont {Kadanoff}(1966)}]{KAD}%
  \BibitemOpen
  \bibfield  {author} {\bibinfo {author} {\bibfnamefont {L.~P.}\ \bibnamefont
  {Kadanoff}},\ }\bibfield  {title} {\bibinfo {title} {Scaling laws for ising
  models near ${T}_{c}$},\ }\href@noop {} {\bibfield  {journal} {\bibinfo
  {journal} {Physics Physique Fizika}\ }\textbf {\bibinfo {volume} {2}},\
  \bibinfo {pages} {263} (\bibinfo {year} {1966})}\BibitemShut {NoStop}%
\bibitem [{\citenamefont {Wetterich}(1991)}]{CWAA}%
  \BibitemOpen
  \bibfield  {author} {\bibinfo {author} {\bibfnamefont {C.}~\bibnamefont
  {Wetterich}},\ }\bibfield  {title} {\bibinfo {title} {Average action and the
  renormalization group equations},\ }\href@noop {} {\bibfield  {journal}
  {\bibinfo  {journal} {Nuclear Physics B}\ }\textbf {\bibinfo {volume}
  {352}},\ \bibinfo {pages} {529} (\bibinfo {year} {1991})}\BibitemShut
  {NoStop}%
\bibitem [{\citenamefont {Wetterich}(1993)}]{CWEFE}%
  \BibitemOpen
  \bibfield  {author} {\bibinfo {author} {\bibfnamefont {C.}~\bibnamefont
  {Wetterich}},\ }\bibfield  {title} {\bibinfo {title} {Exact evolution
  equation for the effective potential},\ }\href@noop {} {\bibfield  {journal}
  {\bibinfo  {journal} {Physics Letters B}\ }\textbf {\bibinfo {volume}
  {301}},\ \bibinfo {pages} {90} (\bibinfo {year} {1993})},\ \Eprint
  {https://arxiv.org/abs/1710.05815} {arXiv:1710.05815 [hep-th]} \BibitemShut
  {NoStop}%
\bibitem [{\citenamefont {Reuter}\ and\ \citenamefont
  {Wetterich}(1994)}]{RWGT}%
  \BibitemOpen
  \bibfield  {author} {\bibinfo {author} {\bibfnamefont {M.}~\bibnamefont
  {Reuter}}\ and\ \bibinfo {author} {\bibfnamefont {C.}~\bibnamefont
  {Wetterich}},\ }\bibfield  {title} {\bibinfo {title} {Effective average
  action for gauge theories and exact evolution equations},\ }\href@noop {}
  {\bibfield  {journal} {\bibinfo  {journal} {Nuclear Physics B}\ }\textbf
  {\bibinfo {volume} {417}},\ \bibinfo {pages} {181} (\bibinfo {year}
  {1994})}\BibitemShut {NoStop}%
\bibitem [{\citenamefont {Tetradis}\ and\ \citenamefont
  {Wetterich}(1994)}]{TETW}%
  \BibitemOpen
  \bibfield  {author} {\bibinfo {author} {\bibfnamefont {N.}~\bibnamefont
  {Tetradis}}\ and\ \bibinfo {author} {\bibfnamefont {C.}~\bibnamefont
  {Wetterich}},\ }\bibfield  {title} {\bibinfo {title} {Critical exponents from
  the effective average action},\ }\href@noop {} {\bibfield  {journal}
  {\bibinfo  {journal} {Nuclear Physics B}\ }\textbf {\bibinfo {volume}
  {422}},\ \bibinfo {pages} {541} (\bibinfo {year} {1994})},\ \Eprint
  {https://arxiv.org/abs/hep-ph/9308214} {arXiv:hep-ph/9308214} \BibitemShut
  {NoStop}%
\bibitem [{\citenamefont {Ellwanger}(1994)}]{ELL}%
  \BibitemOpen
  \bibfield  {author} {\bibinfo {author} {\bibfnamefont {U.}~\bibnamefont
  {Ellwanger}},\ }\bibfield  {title} {\bibinfo {title} {Flow equations forn
  point functions and bound states},\ }\href@noop {} {\bibfield  {journal}
  {\bibinfo  {journal} {Zeitschrift f{\"u}r Physik C Particles and Fields}\
  }\textbf {\bibinfo {volume} {62}},\ \bibinfo {pages} {503–510} (\bibinfo
  {year} {1994})},\ \Eprint {https://arxiv.org/abs/hep-ph/9308260}
  {arXiv:hep-ph/9308260} \BibitemShut {NoStop}%
\bibitem [{\citenamefont {MORRIS}(1994)}]{MOR}%
  \BibitemOpen
  \bibfield  {author} {\bibinfo {author} {\bibfnamefont {T.~R.}\ \bibnamefont
  {MORRIS}},\ }\bibfield  {title} {\bibinfo {title} {The exact renormalization
  group and approximate solutions},\ }\href@noop {} {\bibfield  {journal}
  {\bibinfo  {journal} {International Journal of Modern Physics A}\ }\textbf
  {\bibinfo {volume} {09}},\ \bibinfo {pages} {2411–2449} (\bibinfo {year}
  {1994})},\ \Eprint {https://arxiv.org/abs/hep-ph/9308265}
  {arXiv:hep-ph/9308265} \BibitemShut {NoStop}%
\bibitem [{\citenamefont {Dupuis}\ \emph {et~al.}(2021)\citenamefont {Dupuis},
  \citenamefont {Canet}, \citenamefont {Eichhorn}, \citenamefont {Metzner},
  \citenamefont {Pawlowski}, \citenamefont {Tissier},\ and\ \citenamefont
  {Wschebor}}]{DCE}%
  \BibitemOpen
  \bibfield  {author} {\bibinfo {author} {\bibfnamefont {N.}~\bibnamefont
  {Dupuis}}, \bibinfo {author} {\bibfnamefont {L.}~\bibnamefont {Canet}},
  \bibinfo {author} {\bibfnamefont {A.}~\bibnamefont {Eichhorn}}, \bibinfo
  {author} {\bibfnamefont {W.}~\bibnamefont {Metzner}}, \bibinfo {author}
  {\bibfnamefont {J.}~\bibnamefont {Pawlowski}}, \bibinfo {author}
  {\bibfnamefont {M.}~\bibnamefont {Tissier}},\ and\ \bibinfo {author}
  {\bibfnamefont {N.}~\bibnamefont {Wschebor}},\ }\bibfield  {title} {\bibinfo
  {title} {The nonperturbative functional renormalization group and its
  applications},\ }\href@noop {} {\bibfield  {journal} {\bibinfo  {journal}
  {Physics Reports}\ }\textbf {\bibinfo {volume} {910}},\ \bibinfo {pages} {1}
  (\bibinfo {year} {2021})},\ \Eprint {https://arxiv.org/abs/2006.04853}
  {arXiv:2006.04853 [cond-mat.stat-mech]} \BibitemShut {NoStop}%
\bibitem [{\citenamefont {Ellwanger}\ \emph {et~al.}(1995)\citenamefont
  {Ellwanger}, \citenamefont {Hirsch},\ and\ \citenamefont {Weber}}]{EHW}%
  \BibitemOpen
  \bibfield  {author} {\bibinfo {author} {\bibfnamefont {U.}~\bibnamefont
  {Ellwanger}}, \bibinfo {author} {\bibfnamefont {M.}~\bibnamefont {Hirsch}},\
  and\ \bibinfo {author} {\bibfnamefont {A.}~\bibnamefont {Weber}},\ }\href
  {https://arxiv.org/abs/hep-th/9506019} {\bibinfo {title} {Flow equations for
  the relevant part of the pure yang-mills action}} (\bibinfo {year} {1995}),\
  \Eprint {https://arxiv.org/abs/hep-th/9506019} {arXiv:hep-th/9506019
  [hep-th]} \BibitemShut {NoStop}%
\bibitem [{\citenamefont {Freire}\ and\ \citenamefont {Wetterich}(1996)}]{FRW}%
  \BibitemOpen
  \bibfield  {author} {\bibinfo {author} {\bibfnamefont {F.}~\bibnamefont
  {Freire}}\ and\ \bibinfo {author} {\bibfnamefont {C.}~\bibnamefont
  {Wetterich}},\ }\bibfield  {title} {\bibinfo {title} {Abelian ward identity
  from the background field dependence of the effective action},\ }\href@noop
  {} {\bibfield  {journal} {\bibinfo  {journal} {Physics Letters B}\ }\textbf
  {\bibinfo {volume} {380}},\ \bibinfo {pages} {337} (\bibinfo {year}
  {1996})},\ \Eprint {https://arxiv.org/abs/hep-th/9601081}
  {arXiv:hep-th/9601081} \BibitemShut {NoStop}%
\bibitem [{\citenamefont {Freire}\ \emph {et~al.}(2000)\citenamefont {Freire},
  \citenamefont {Litim},\ and\ \citenamefont {Pawlowski}}]{FDP}%
  \BibitemOpen
  \bibfield  {author} {\bibinfo {author} {\bibfnamefont {F.}~\bibnamefont
  {Freire}}, \bibinfo {author} {\bibfnamefont {D.~F.}\ \bibnamefont {Litim}},\
  and\ \bibinfo {author} {\bibfnamefont {J.~M.}\ \bibnamefont {Pawlowski}},\
  }\bibfield  {title} {\bibinfo {title} {Gauge invariance and background field
  formalism in the exact renormalisation group},\ }\href@noop {} {\bibfield
  {journal} {\bibinfo  {journal} {Physics Letters B}\ }\textbf {\bibinfo
  {volume} {495}},\ \bibinfo {pages} {256} (\bibinfo {year} {2000})},\ \Eprint
  {https://arxiv.org/abs/hep-th/0009110} {arXiv:hep-th/0009110} \BibitemShut
  {NoStop}%
\bibitem [{\citenamefont {Wetterich}(2015)}]{Wetterich_2015}%
  \BibitemOpen
  \bibfield  {author} {\bibinfo {author} {\bibfnamefont {C.}~\bibnamefont
  {Wetterich}},\ }\bibfield  {title} {\bibinfo {title} {Cosmic fluctuations
  from a quantum effective action},\ }\href@noop {} {\bibfield  {journal}
  {\bibinfo  {journal} {Physical Review D}\ }\textbf {\bibinfo {volume} {92}}
  (\bibinfo {year} {2015})}\BibitemShut {NoStop}%
\bibitem [{\citenamefont {Wetterich}(2017)}]{Wetterich_2017}%
  \BibitemOpen
  \bibfield  {author} {\bibinfo {author} {\bibfnamefont {C.}~\bibnamefont
  {Wetterich}},\ }\bibfield  {title} {\bibinfo {title} {Quantum correlations
  for the metric},\ }\href@noop {} {\bibfield  {journal} {\bibinfo  {journal}
  {Physical Review D}\ }\textbf {\bibinfo {volume} {95}} (\bibinfo {year}
  {2017})}\BibitemShut {NoStop}%
\bibitem [{\citenamefont {Floerchinger}(2012)}]{FLO}%
  \BibitemOpen
  \bibfield  {author} {\bibinfo {author} {\bibfnamefont {S.}~\bibnamefont
  {Floerchinger}},\ }\bibfield  {title} {\bibinfo {title} {Analytic
  continuation of functional renormalization group equations},\ }\href@noop {}
  {\bibfield  {journal} {\bibinfo  {journal} {Journal of High Energy Physics}\
  }\textbf {\bibinfo {volume} {2012}} (\bibinfo {year} {2012})},\ \Eprint
  {https://arxiv.org/abs/1112.4374} {arXiv:1112.4374 [hep-th]} \BibitemShut
  {NoStop}%
\bibitem [{\citenamefont {Pawlowski}\ and\ \citenamefont
  {Reichert}(2021)}]{PARE}%
  \BibitemOpen
  \bibfield  {author} {\bibinfo {author} {\bibfnamefont {J.~M.}\ \bibnamefont
  {Pawlowski}}\ and\ \bibinfo {author} {\bibfnamefont {M.}~\bibnamefont
  {Reichert}},\ }\bibfield  {title} {\bibinfo {title} {Quantum gravity: A
  fluctuating point of view},\ }\href@noop {} {\bibfield  {journal} {\bibinfo
  {journal} {Frontiers in Physics}\ }\textbf {\bibinfo {volume} {8}} (\bibinfo
  {year} {2021})},\ \Eprint {https://arxiv.org/abs/2007.10353}
  {arXiv:2007.10353 [hep-th]} \BibitemShut {NoStop}%
\bibitem [{\citenamefont {Fehre}\ \emph {et~al.}(2023)\citenamefont {Fehre},
  \citenamefont {Litim}, \citenamefont {Pawlowski},\ and\ \citenamefont
  {Reichert}}]{FLPR}%
  \BibitemOpen
  \bibfield  {author} {\bibinfo {author} {\bibfnamefont {J.}~\bibnamefont
  {Fehre}}, \bibinfo {author} {\bibfnamefont {D.~F.}\ \bibnamefont {Litim}},
  \bibinfo {author} {\bibfnamefont {J.~M.}\ \bibnamefont {Pawlowski}},\ and\
  \bibinfo {author} {\bibfnamefont {M.}~\bibnamefont {Reichert}},\ }\href@noop
  {} {\bibinfo {title} {Lorentzian quantum gravity and the graviton spectral
  function}} (\bibinfo {year} {2023}),\ \Eprint
  {https://arxiv.org/abs/2111.13232} {arXiv:2111.13232 [hep-th]} \BibitemShut
  {NoStop}%
\bibitem [{\citenamefont {Braun}\ \emph {et~al.}(2023)\citenamefont {Braun},
  \citenamefont {rui Chen}, \citenamefont {jie Fu}, \citenamefont {Geißel},
  \citenamefont {Horak}, \citenamefont {Huang}, \citenamefont {Ihssen},
  \citenamefont {Pawlowski}, \citenamefont {Reichert}, \citenamefont
  {Rennecke}, \citenamefont {yang Tan}, \citenamefont {Töpfel}, \citenamefont
  {Wessely},\ and\ \citenamefont {Wink}}]{JB}%
  \BibitemOpen
  \bibfield  {author} {\bibinfo {author} {\bibfnamefont {J.}~\bibnamefont
  {Braun}}, \bibinfo {author} {\bibfnamefont {Y.}~\bibnamefont {rui Chen}},
  \bibinfo {author} {\bibfnamefont {W.}~\bibnamefont {jie Fu}}, \bibinfo
  {author} {\bibfnamefont {A.}~\bibnamefont {Geißel}}, \bibinfo {author}
  {\bibfnamefont {J.}~\bibnamefont {Horak}}, \bibinfo {author} {\bibfnamefont
  {C.}~\bibnamefont {Huang}}, \bibinfo {author} {\bibfnamefont
  {F.}~\bibnamefont {Ihssen}}, \bibinfo {author} {\bibfnamefont {J.~M.}\
  \bibnamefont {Pawlowski}}, \bibinfo {author} {\bibfnamefont {M.}~\bibnamefont
  {Reichert}}, \bibinfo {author} {\bibfnamefont {F.}~\bibnamefont {Rennecke}},
  \bibinfo {author} {\bibfnamefont {Y.}~\bibnamefont {yang Tan}}, \bibinfo
  {author} {\bibfnamefont {S.}~\bibnamefont {Töpfel}}, \bibinfo {author}
  {\bibfnamefont {J.}~\bibnamefont {Wessely}},\ and\ \bibinfo {author}
  {\bibfnamefont {N.}~\bibnamefont {Wink}},\ }\href@noop {} {\bibinfo {title}
  {Renormalised spectral flows}} (\bibinfo {year} {2023}),\ \Eprint
  {https://arxiv.org/abs/2206.10232} {arXiv:2206.10232 [hep-th]} \BibitemShut
  {NoStop}%
\bibitem [{\citenamefont {Pawlowski}\ and\ \citenamefont
  {Reichert}(2023)}]{PARE2}%
  \BibitemOpen
  \bibfield  {author} {\bibinfo {author} {\bibfnamefont {J.~M.}\ \bibnamefont
  {Pawlowski}}\ and\ \bibinfo {author} {\bibfnamefont {M.}~\bibnamefont
  {Reichert}},\ }\href@noop {} {\bibinfo {title} {Quantum gravity from
  dynamical metric fluctuations}} (\bibinfo {year} {2023}),\ \Eprint
  {https://arxiv.org/abs/2309.10785} {arXiv:2309.10785 [hep-th]} \BibitemShut
  {NoStop}%
\bibitem [{\citenamefont {Litim}(2001)}]{LIT}%
  \BibitemOpen
  \bibfield  {author} {\bibinfo {author} {\bibfnamefont {D.~F.}\ \bibnamefont
  {Litim}},\ }\bibfield  {title} {\bibinfo {title} {Optimized renormalization
  group flows},\ }\href@noop {} {\bibfield  {journal} {\bibinfo  {journal}
  {Physical Review D}\ }\textbf {\bibinfo {volume} {64}} (\bibinfo {year}
  {2001})},\ \Eprint {https://arxiv.org/abs/hep-th/0103195}
  {arXiv:hep-th/0103195} \BibitemShut {NoStop}%
\bibitem [{\citenamefont {Wetterich}(2016)}]{CWGIFE}%
  \BibitemOpen
  \bibfield  {author} {\bibinfo {author} {\bibfnamefont {C.}~\bibnamefont
  {Wetterich}},\ }\href@noop {} {\bibinfo {title} {Gauge invariant flow
  equation}} (\bibinfo {year} {2016}),\ \Eprint
  {https://arxiv.org/abs/1607.02989} {arXiv:1607.02989 [hep-th]} \BibitemShut
  {NoStop}%
\bibitem [{\citenamefont {Wetterich}(2021)}]{CWFSI}%
  \BibitemOpen
  \bibfield  {author} {\bibinfo {author} {\bibfnamefont {C.}~\bibnamefont
  {Wetterich}},\ }\bibfield  {title} {\bibinfo {title} {Fundamental scale
  invariance},\ }\href@noop {} {\bibfield  {journal} {\bibinfo  {journal}
  {Nuclear Physics B}\ }\textbf {\bibinfo {volume} {964}},\ \bibinfo {pages}
  {115326} (\bibinfo {year} {2021})},\ \Eprint
  {https://arxiv.org/abs/2007.08805} {arXiv:2007.08805 [hep-th]} \BibitemShut
  {NoStop}%
\bibitem [{\citenamefont {Wetterich}\ and\ \citenamefont
  {Yamada}(2019)}]{CWYA}%
  \BibitemOpen
  \bibfield  {author} {\bibinfo {author} {\bibfnamefont {C.}~\bibnamefont
  {Wetterich}}\ and\ \bibinfo {author} {\bibfnamefont {M.}~\bibnamefont
  {Yamada}},\ }\bibfield  {title} {\bibinfo {title} {Variable planck mass from
  the gauge invariant flow equation},\ }\href@noop {} {\bibfield  {journal}
  {\bibinfo  {journal} {Physical Review D}\ }\textbf {\bibinfo {volume} {100}}
  (\bibinfo {year} {2019})},\ \Eprint {https://arxiv.org/abs/1811.11706}
  {arXiv:1811.11706 [hep-th]} \BibitemShut {NoStop}%
\bibitem [{\citenamefont {Oleszczuk}(1994)}]{OLE}%
  \BibitemOpen
  \bibfield  {author} {\bibinfo {author} {\bibfnamefont {M.}~\bibnamefont
  {Oleszczuk}},\ }\bibfield  {title} {\bibinfo {title} {A symmetry-preserving
  cut-off regularization},\ }\href@noop {} {\bibfield  {journal} {\bibinfo
  {journal} {Zeitschrift f{\"u}r Physik C Particles and Fields}\ }\textbf
  {\bibinfo {volume} {64}},\ \bibinfo {pages} {533} (\bibinfo {year}
  {1994})}\BibitemShut {NoStop}%
\bibitem [{\citenamefont {Floreanini}\ and\ \citenamefont
  {Percacci}(1995)}]{FP}%
  \BibitemOpen
  \bibfield  {author} {\bibinfo {author} {\bibfnamefont {R.}~\bibnamefont
  {Floreanini}}\ and\ \bibinfo {author} {\bibfnamefont {R.}~\bibnamefont
  {Percacci}},\ }\bibfield  {title} {\bibinfo {title} {The heat-kernel and the
  average effective potential},\ }\href@noop {} {\bibfield  {journal} {\bibinfo
   {journal} {Physics Letters B}\ }\textbf {\bibinfo {volume} {356}},\ \bibinfo
  {pages} {205–210} (\bibinfo {year} {1995})},\ \Eprint
  {https://arxiv.org/abs/hep-th/9505172} {arXiv:hep-th/9505172} \BibitemShut
  {NoStop}%
\bibitem [{\citenamefont {Liao}(1996)}]{LI}%
  \BibitemOpen
  \bibfield  {author} {\bibinfo {author} {\bibfnamefont {S.-B.}\ \bibnamefont
  {Liao}},\ }\bibfield  {title} {\bibinfo {title} {Connection between momentum
  cutoff and operator cutoff regularizations},\ }\href@noop {} {\bibfield
  {journal} {\bibinfo  {journal} {Physical Review D}\ }\textbf {\bibinfo
  {volume} {53}},\ \bibinfo {pages} {2020–2036} (\bibinfo {year} {1996})},\
  \Eprint {https://arxiv.org/abs/hep-th/9501124} {arXiv:hep-th/9501124}
  \BibitemShut {NoStop}%
\bibitem [{\citenamefont {Liao}(1997)}]{LIA}%
  \BibitemOpen
  \bibfield  {author} {\bibinfo {author} {\bibfnamefont {S.-B.}\ \bibnamefont
  {Liao}},\ }\bibfield  {title} {\bibinfo {title} {Operator cutoff
  regularization and renormalization group in yang-mills theory},\ }\href@noop
  {} {\bibfield  {journal} {\bibinfo  {journal} {Physical Review D}\ }\textbf
  {\bibinfo {volume} {56}},\ \bibinfo {pages} {5008–5033} (\bibinfo {year}
  {1997})},\ \Eprint {https://arxiv.org/abs/hep-th/9511046}
  {arXiv:hep-th/9511046} \BibitemShut {NoStop}%
\bibitem [{\citenamefont {Schaefer}\ and\ \citenamefont {Pirner}(1997)}]{SP}%
  \BibitemOpen
  \bibfield  {author} {\bibinfo {author} {\bibfnamefont {B.~J.}\ \bibnamefont
  {Schaefer}}\ and\ \bibinfo {author} {\bibfnamefont {H.~J.}\ \bibnamefont
  {Pirner}},\ }\href@noop {} {\bibinfo {title} {Nonperturbative flow equations
  with heat-kernel methods at finite temperature}} (\bibinfo {year} {1997}),\
  \Eprint {https://arxiv.org/abs/hep-ph/9712413} {arXiv:hep-ph/9712413}
  \BibitemShut {NoStop}%
\bibitem [{\citenamefont {Bonanno}\ and\ \citenamefont {Zappalà}(2001)}]{BOZ}%
  \BibitemOpen
  \bibfield  {author} {\bibinfo {author} {\bibfnamefont {A.}~\bibnamefont
  {Bonanno}}\ and\ \bibinfo {author} {\bibfnamefont {D.}~\bibnamefont
  {Zappalà}},\ }\bibfield  {title} {\bibinfo {title} {Towards an accurate
  determination of the critical exponents with the renormalization group flow
  equations},\ }\href@noop {} {\bibfield  {journal} {\bibinfo  {journal}
  {Physics Letters B}\ }\textbf {\bibinfo {volume} {504}},\ \bibinfo {pages}
  {181–187} (\bibinfo {year} {2001})},\ \Eprint
  {https://arxiv.org/abs/hep-th/0010095} {arXiv:hep-th/0010095} \BibitemShut
  {NoStop}%
\bibitem [{\citenamefont {BOHR}\ \emph {et~al.}(2001)\citenamefont {BOHR},
  \citenamefont {SCHAEFER},\ and\ \citenamefont {WAMBACH}}]{BSW}%
  \BibitemOpen
  \bibfield  {author} {\bibinfo {author} {\bibfnamefont {O.}~\bibnamefont
  {BOHR}}, \bibinfo {author} {\bibfnamefont {B.-J.}\ \bibnamefont {SCHAEFER}},\
  and\ \bibinfo {author} {\bibfnamefont {J.}~\bibnamefont {WAMBACH}},\
  }\bibfield  {title} {\bibinfo {title} {Renormalization group flow equations
  and the phase transition in o(n)-models},\ }\href@noop {} {\bibfield
  {journal} {\bibinfo  {journal} {International Journal of Modern Physics A}\
  }\textbf {\bibinfo {volume} {16}},\ \bibinfo {pages} {3823–3852} (\bibinfo
  {year} {2001})},\ \Eprint {https://arxiv.org/abs/hep-ph/0007098}
  {arXiv:hep-ph/0007098} \BibitemShut {NoStop}%
\bibitem [{\citenamefont {Zappalà}(2001)}]{ZAP}%
  \BibitemOpen
  \bibfield  {author} {\bibinfo {author} {\bibfnamefont {D.}~\bibnamefont
  {Zappalà}},\ }\bibfield  {title} {\bibinfo {title} {Improving the
  renormalization group approach to the quantum-mechanical double well
  potential},\ }\href@noop {} {\bibfield  {journal} {\bibinfo  {journal}
  {Physics Letters A}\ }\textbf {\bibinfo {volume} {290}},\ \bibinfo {pages}
  {35–40} (\bibinfo {year} {2001})},\ \Eprint
  {https://arxiv.org/abs/quant-ph/0108019} {arXiv:quant-ph/0108019}
  \BibitemShut {NoStop}%
\bibitem [{\citenamefont {Litim}\ and\ \citenamefont
  {Pawlowski}(2001)}]{LIPA1}%
  \BibitemOpen
  \bibfield  {author} {\bibinfo {author} {\bibfnamefont {D.~F.}\ \bibnamefont
  {Litim}}\ and\ \bibinfo {author} {\bibfnamefont {J.~M.}\ \bibnamefont
  {Pawlowski}},\ }\bibfield  {title} {\bibinfo {title} {Predictive power of
  renormalisation group flows: a comparison},\ }\href@noop {} {\bibfield
  {journal} {\bibinfo  {journal} {Physics Letters B}\ }\textbf {\bibinfo
  {volume} {516}},\ \bibinfo {pages} {197–207} (\bibinfo {year} {2001})},\
  \Eprint {https://arxiv.org/abs/hep-th/0107020} {arXiv:hep-th/0107020}
  \BibitemShut {NoStop}%
\bibitem [{\citenamefont {Litim}\ and\ \citenamefont
  {Pawlowski}(2002{\natexlab{a}})}]{LIPA2}%
  \BibitemOpen
  \bibfield  {author} {\bibinfo {author} {\bibfnamefont {D.~F.}\ \bibnamefont
  {Litim}}\ and\ \bibinfo {author} {\bibfnamefont {J.~M.}\ \bibnamefont
  {Pawlowski}},\ }\bibfield  {title} {\bibinfo {title} {Perturbation theory and
  renormalization group equations},\ }\href@noop {} {\bibfield  {journal}
  {\bibinfo  {journal} {Physical Review D}\ }\textbf {\bibinfo {volume} {65}}
  (\bibinfo {year} {2002}{\natexlab{a}})},\ \Eprint
  {https://arxiv.org/abs/hep-th/0111191} {arXiv:hep-th/0111191} \BibitemShut
  {NoStop}%
\bibitem [{\citenamefont {Zappalà}(2002)}]{ZAP2}%
  \BibitemOpen
  \bibfield  {author} {\bibinfo {author} {\bibfnamefont {D.}~\bibnamefont
  {Zappalà}},\ }\bibfield  {title} {\bibinfo {title} {Perturbative and
  nonperturbative aspects of the proper time renormalization group},\
  }\href@noop {} {\bibfield  {journal} {\bibinfo  {journal} {Physical Review
  D}\ }\textbf {\bibinfo {volume} {66}} (\bibinfo {year} {2002})},\ \Eprint
  {https://arxiv.org/abs/hep-th/0202167} {arXiv:hep-th/0202167} \BibitemShut
  {NoStop}%
\bibitem [{\citenamefont {Bonanno}\ and\ \citenamefont {Reuter}(2005)}]{BR}%
  \BibitemOpen
  \bibfield  {author} {\bibinfo {author} {\bibfnamefont {A.}~\bibnamefont
  {Bonanno}}\ and\ \bibinfo {author} {\bibfnamefont {M.}~\bibnamefont
  {Reuter}},\ }\bibfield  {title} {\bibinfo {title} {Proper time flow equation
  for gravity},\ }\href@noop {} {\bibfield  {journal} {\bibinfo  {journal}
  {Journal of High Energy Physics}\ }\textbf {\bibinfo {volume} {2005}},\
  \bibinfo {pages} {035–035} (\bibinfo {year} {2005})},\ \Eprint
  {https://arxiv.org/abs/hep-th/0410191} {arXiv:hep-th/0410191} \BibitemShut
  {NoStop}%
\bibitem [{\citenamefont {de~Alwis}(2018)}]{DA}%
  \BibitemOpen
  \bibfield  {author} {\bibinfo {author} {\bibfnamefont {S.}~\bibnamefont
  {de~Alwis}},\ }\bibfield  {title} {\bibinfo {title} {Exact rg flow equations
  and quantum gravity},\ }\href@noop {} {\bibfield  {journal} {\bibinfo
  {journal} {Journal of High Energy Physics}\ }\textbf {\bibinfo {volume}
  {2018}} (\bibinfo {year} {2018})},\ \Eprint
  {https://arxiv.org/abs/1707.09298} {arXiv:1707.09298 [hep-th]} \BibitemShut
  {NoStop}%
\bibitem [{\citenamefont {Abel}\ and\ \citenamefont {Heurtier}(2023)}]{AH}%
  \BibitemOpen
  \bibfield  {author} {\bibinfo {author} {\bibfnamefont {S.}~\bibnamefont
  {Abel}}\ and\ \bibinfo {author} {\bibfnamefont {L.}~\bibnamefont
  {Heurtier}},\ }\href@noop {} {\bibinfo {title} {Exact schwinger proper time
  renormalisation}} (\bibinfo {year} {2023}),\ \Eprint
  {https://arxiv.org/abs/2311.12102} {arXiv:2311.12102 [hep-th]} \BibitemShut
  {NoStop}%
\bibitem [{\citenamefont {Bonanno}\ \emph {et~al.}(2020)\citenamefont
  {Bonanno}, \citenamefont {Lippoldt}, \citenamefont {Percacci},\ and\
  \citenamefont {Vacca}}]{BLPV}%
  \BibitemOpen
  \bibfield  {author} {\bibinfo {author} {\bibfnamefont {A.}~\bibnamefont
  {Bonanno}}, \bibinfo {author} {\bibfnamefont {S.}~\bibnamefont {Lippoldt}},
  \bibinfo {author} {\bibfnamefont {R.}~\bibnamefont {Percacci}},\ and\
  \bibinfo {author} {\bibfnamefont {G.~P.}\ \bibnamefont {Vacca}},\ }\bibfield
  {title} {\bibinfo {title} {On exact proper time wilsonian rg flows},\
  }\href@noop {} {\bibfield  {journal} {\bibinfo  {journal} {The European
  Physical Journal C}\ }\textbf {\bibinfo {volume} {80}} (\bibinfo {year}
  {2020})},\ \Eprint {https://arxiv.org/abs/1912.08135} {arXiv:1912.08135
  [hep-th]} \BibitemShut {NoStop}%
\bibitem [{\citenamefont {Litim}\ and\ \citenamefont
  {Pawlowski}(2002{\natexlab{b}})}]{Litim_2002}%
  \BibitemOpen
  \bibfield  {author} {\bibinfo {author} {\bibfnamefont {D.~F.}\ \bibnamefont
  {Litim}}\ and\ \bibinfo {author} {\bibfnamefont {J.~M.}\ \bibnamefont
  {Pawlowski}},\ }\bibfield  {title} {\bibinfo {title} {Completeness and
  consistency of renormalization group flows},\ }\href@noop {} {\bibfield
  {journal} {\bibinfo  {journal} {Physical Review D}\ }\textbf {\bibinfo
  {volume} {66}} (\bibinfo {year} {2002}{\natexlab{b}})},\ \Eprint
  {https://arxiv.org/abs/hep-ph/0202188} {arXiv:hep-ph/0202188} \BibitemShut
  {NoStop}%
\bibitem [{\citenamefont {Litim}\ and\ \citenamefont
  {Pawlowski}(2002{\natexlab{c}})}]{Litim_2002B}%
  \BibitemOpen
  \bibfield  {author} {\bibinfo {author} {\bibfnamefont {D.~F.}\ \bibnamefont
  {Litim}}\ and\ \bibinfo {author} {\bibfnamefont {J.~M.}\ \bibnamefont
  {Pawlowski}},\ }\bibfield  {title} {\bibinfo {title} {Wilsonian flows and
  background fields},\ }\href {https://doi.org/10.1016/s0370-2693(02)02693-x}
  {\bibfield  {journal} {\bibinfo  {journal} {Physics Letters B}\ }\textbf
  {\bibinfo {volume} {546}},\ \bibinfo {pages} {279–286} (\bibinfo {year}
  {2002}{\natexlab{c}})},\ \Eprint {https://arxiv.org/abs/hep-ph/0208216}
  {arXiv:hep-ph/0208216} \BibitemShut {NoStop}%
\bibitem [{\citenamefont {Pawlowski}\ \emph {et~al.}(2019)\citenamefont
  {Pawlowski}, \citenamefont {Reichert}, \citenamefont {Wetterich},\ and\
  \citenamefont {Yamada}}]{PRWY}%
  \BibitemOpen
  \bibfield  {author} {\bibinfo {author} {\bibfnamefont {J.~M.}\ \bibnamefont
  {Pawlowski}}, \bibinfo {author} {\bibfnamefont {M.}~\bibnamefont {Reichert}},
  \bibinfo {author} {\bibfnamefont {C.}~\bibnamefont {Wetterich}},\ and\
  \bibinfo {author} {\bibfnamefont {M.}~\bibnamefont {Yamada}},\ }\bibfield
  {title} {\bibinfo {title} {Higgs scalar potential in asymptotically safe
  quantum gravity},\ }\href@noop {} {\bibfield  {journal} {\bibinfo  {journal}
  {Physical Review D}\ }\textbf {\bibinfo {volume} {99}} (\bibinfo {year}
  {2019})},\ \Eprint {https://arxiv.org/abs/1811.11706} {arXiv:1811.11706
  [hep-th]} \BibitemShut {NoStop}%
\bibitem [{\citenamefont {Wetterich}(2019)}]{ESPA}%
  \BibitemOpen
  \bibfield  {author} {\bibinfo {author} {\bibfnamefont {C.}~\bibnamefont
  {Wetterich}},\ }\href@noop {} {\bibinfo {title} {Effective scalar potential
  in asymptotically safe quantum gravity}} (\bibinfo {year} {2019}),\ \Eprint
  {https://arxiv.org/abs/1911.06100} {arXiv:1911.06100 [hep-th]} \BibitemShut
  {NoStop}%
\bibitem [{\citenamefont {Wetterich}(2022)}]{CWQGS}%
  \BibitemOpen
  \bibfield  {author} {\bibinfo {author} {\bibfnamefont {C.}~\bibnamefont
  {Wetterich}},\ }\href@noop {} {\bibinfo {title} {Quantum gravity and scale
  symmetry in cosmology}} (\bibinfo {year} {2022}),\ \Eprint
  {https://arxiv.org/abs/2211.03596} {arXiv:2211.03596 [gr-qc]} \BibitemShut
  {NoStop}%
\bibitem [{\citenamefont {Papenbrock}\ and\ \citenamefont
  {Wetterich}(1995)}]{Papenbrock_1995}%
  \BibitemOpen
  \bibfield  {author} {\bibinfo {author} {\bibfnamefont {T.}~\bibnamefont
  {Papenbrock}}\ and\ \bibinfo {author} {\bibfnamefont {C.}~\bibnamefont
  {Wetterich}},\ }\bibfield  {title} {\bibinfo {title} {Two-loop results from
  improved one loop computations},\ }\href@noop {} {\bibfield  {journal}
  {\bibinfo  {journal} {Zeitschrift für Physik C Particles and Fields}\
  }\textbf {\bibinfo {volume} {65}},\ \bibinfo {pages} {519–535} (\bibinfo
  {year} {1995})}\BibitemShut {NoStop}%
\bibitem [{\citenamefont {Wetterich}(2024)}]{wetterich2024field}%
  \BibitemOpen
  \bibfield  {author} {\bibinfo {author} {\bibfnamefont {C.}~\bibnamefont
  {Wetterich}},\ }\href@noop {} {\bibinfo {title} {Field transformations in
  functional integral, effective action and functional flow equations}}
  (\bibinfo {year} {2024}),\ \Eprint {https://arxiv.org/abs/2402.04679}
  {arXiv:2402.04679 [hep-th]} \BibitemShut {NoStop}%
\bibitem [{\citenamefont {Wetterich}(2018)}]{CWGIYM}%
  \BibitemOpen
  \bibfield  {author} {\bibinfo {author} {\bibfnamefont {C.}~\bibnamefont
  {Wetterich}},\ }\bibfield  {title} {\bibinfo {title} {Gauge-invariant fields
  and flow equations for yang–mills theories},\ }\href@noop {} {\bibfield
  {journal} {\bibinfo  {journal} {Nuclear Physics B}\ }\textbf {\bibinfo
  {volume} {934}},\ \bibinfo {pages} {265–316} (\bibinfo {year} {2018})},\
  \Eprint {https://arxiv.org/abs/1710.02494} {arXiv:1710.02494 [hep-th]}
  \BibitemShut {NoStop}%
\end{thebibliography}%

\end{document}